\documentclass{revtex4-2}
\usepackage{amsmath}
\usepackage{bm}
\usepackage{amssymb}
\usepackage{mathtools}
\usepackage{color}
\usepackage{graphicx}
\usepackage{acronym,xspace}
\usepackage{macros}
\usepackage{array, multirow}   
\usepackage{booktabs}
\usepackage{tabularx}
\usepackage{geometry}
\usepackage{pythonhighlight}

\usepackage{hyperref}
\hypersetup{
    colorlinks=true,
    linkcolor=blue,
    filecolor=magenta,      
    urlcolor=cyan,
}

\begin{document}

\title[LIGOFRACTALS]{Characterization of gravitational-wave detector noise with fractals}

\author{Marco Cavagli\`a}
\affiliation{Institute of Multi-messenger Astrophysics and Cosmology, Missouri University of Science and Technology, Physics Building, 1315 N.\ Pine St., Rolla, MO 65409, USA}

\begin{abstract}
We present a new method, based on fractal analysis, to characterize the output of a physical detector that is in the form of a set of real-valued, discrete physical measurements. We apply the method to gravitational-wave data from the latest observing run of the Laser Interferometer Gravitational-wave Observatory. We show that a measure of the fractal dimension of the main detector output (strain channel) can be used to determine the instrument status, test data stationarity, and identify non-astrophysical excess noise in low latency. When applied to instrument control and environmental data (auxiliary channels) the fractal dimension can be used to identify the origins of noise transients, non-linear couplings in the various detector subsystems, and provide a means to flag stretches of low-quality data.
\end{abstract}

\maketitle

\section{Introduction}\label{introduction}

This paper introduces a new method to characterize the data of \ac{GW} interferometric detectors and, more generally, the data of any detector whose output is in the form of a (set of) real-valued, discretely-sampled measurements. Current ground-based \ac{GW} detectors such as \ac{LIGO} \cite{advLigo2015}, Virgo \cite{advVirgo2015}, and KAGRA \cite{KAGRA:2020tym}, are exquisite scientific instruments of extreme sensitivity. Because of this feature and the feebleness of the signals these detectors aim to detect, coveted astrophysical information is typically buried in instrumental and environmental noise of different origins. Therefore, \ac{GW} scientists devote considerable effort to the characterization of detector noise and data quality investigations \cite{Davis_2021,LIGOScientific:2016gtq,LIGOScientific:2019hgc,VIRGO:2012oxz,LIGOScientific:2014qfs,Slutsky:2010ff}. These activities are essential to improve the instruments, provide feedback to commissioners, validate detection candidates, reduce search backgrounds, and ultimately increase the significance of \ac{GW} signals. 

The noise floor of \ac{GW} detectors is typically non-stationary and non-Gaussian \cite{LIGOScientific:2019hgc}. The detector sensitivity is limited by fundamental and technical noise sources, as
well as transient and persistent noise artifacts that arise from physical disturbances and/or non-linear couplings between the various detector subsystems and their environments \cite{Davis_2021}.
Therefore, characterization of \ac{GW} detector noise is a non-trivial task. To complicate the matter, during observing runs there is a strong need to perform detector characterization and data
quality assessment in low-latency \cite{LValerts}. A rapid and accurate estimation of detector data is indeed crucial to validate or retract candidate detections and disseminate accurate physical
information of astrophysical triggers. Online and offline validation of \ac{GW} triggers requires a series of steps to assess the state of the detector, evaluate specific metrics, and ultimately
produce a single event validation result \cite{LIGOScientific:2019hgc}. Great advances have been made in the automation of online and offline validation of triggers that minimize human involvement
and shorten the latency of the process. Tasks that are currently implemented in \ac{LVK} low-latency data quality reports are based on techniques that include signal processing tools such as
omicron \cite{Robinet:2020lbf}, machine learning algorithms such as iDQ \cite{Essick:2020}, automated checks of lock status and noise stationarity, and monitors to identify physical couplings
between the detectors and their environments. Nevertheless, many of these process still require a human input for final validation. Although identifying noise artifacts with spectrograms and
available auxiliary channel information is relatively easy, assessing their impact on astrophysical candidate trigger parameters and sky locations is not trivial
\cite{Powell:2018csz,Chatziioannou:2021ezd}. Even questions that are in theory as simple as determining whether the detectors are in a good state may be hard to answer. In the following we will
show that a measurement of the fractal dimension of the detector data may be provide an additional, effective tool for characterizing the instrument output in low latency and providing data quality
assessment. 

Fractals arise naturally in non-linear dynamical systems and have been used to study physical problems in a plethora of disciplines including, among others, biology and medicine
\cite{nonnenmacher2013fractals}, engineering \cite{levy2005fractals}, computer graphics \cite{encarnacao2012fractal}, archeology \cite{Brown2005TheBP}, and economics \cite{peters1994fractal}.  In
the context of \ac{GW} interferometry, they have been used to characterize the time evolution of Virgo and KAGRA seismometer data \cite{Bianchi2020, longo2020, longo2020-2, longo2021}. Fractal
sets can be characterized by their dimension. The fractal dimension $\fdim$ of a set determines its degree of complexity \cite{Mandelbrot1977FractalGO}. In the case of a physical measurement
over a time interval, it has been shown that the value of the fractal dimension is related to the frequency content of the data \cite{1990PhyD...46..254H}. For example, Gaussian (white) noise has
$\fdim=2$, while noise with a different power spectrum has typically $1<\fdim<2$. Therefore, the fractal dimension can be used to measure the characteristics of the instrument's noise, as well as
to monitor its stationarity over time. As the fractal dimension of a device's output with stationary noise floor is constant during normal operations, any change in the fractal dimension must
denote a variation in the noise power spectrum of the instrument. The fractal dimension is also particularly suited to capture non-linear effects in complex physical devices.

In the case at hand, this paper will show that fractal analysis can be used to characterize the behavior of \ac{GW} detectors and identify data non-stationarity in their output. A simple measure of
the fractal dimension of the main detector output (the ``strain channel'') enables fast identification of the interferometer lock status, instrumental or environmental excess noise in data used
for astrophysical searches, and also monitors the stationarity of the detector \cite{g2netseminar,g2netschool,ipam}. When applied to instrument control and environmental data (``auxiliary
channels'') the fractal dimension can be used to identify the origins of noise transients, non-linear couplings in the various detector subsystems, and provide a means to flag stretches of
low-quality data. Moreover, the numerical calculation of (a good approximant to) the fractal dimension is computationally cheap and can easily be performed in real time, thus enabling on-the-fly
information on the instrument output \cite{g2netseminar,g2netschool,ipam}. 

For the sake of brevity, the following analysis will focus on LIGO data. Extensions of this analysis to other \ac{GW} detector data, as well as discretely-sampled, real-valued data produced by generic measurement devices are straightforward.

\section{Fractal dimension}\label{FD}

A detailed discussion of the mathematics of fractals is beyond the scope of this paper. The content of this section is limited to a brief discussion of the basic elements of fractal analysis that are required to understand its application to \ac{GW} detector data. 

In broad terms, fractals are subsets of $n$-dimensional Euclidean spaces with non-integer dimension $\fdim<n$ \cite{Mandelbrot1977FractalGO}. The fractal dimension defines the degree of complexity
of the set and quantifies how densely the fractal set occupies its covering $n$-dimensional Euclidean space. Different definitions of fractal dimensions exist in the literature, such as the
Hausdorff dimension  \cite{10.1214/11-STS370}, the packing dimension \cite{tricot_1982}, and the box-counting dimension \cite{Bouligand,MinkowskiBouligand}. These different definitions are generally
equivalent for exactly self-similar fractal sets, but not for generic fractals. Moreover, discretely-sampled physical measurements do not strictly define a fractal set and allow only for an
approximate measure of $\fdim$. Although the fractal dimension of a physical measurement cannot be uniquely defined, the absolute value of $\fdim$ is unimportant for the purpose of this study; all
the {\em relevant} information for the characterization of the instrument noise is encoded in the {\em variation} of $\fdim$. Therefore, throughout this paper we will use the box-counting
(Minkowski-Bouligand) definition of fractal dimension \cite{Bouligand,MinkowskiBouligand} without loss of generality.

Consider a set of real-valued, time-dependent measurements ${\mathcal M}=\{t,{\mathcal O}(t)\}$, where $t\in[0,T]$ and ${\mathcal O}:[0,T]\to\mathbb{R}$. The set ${\mathcal M}$ defines our (fractal) curve. Let ${\mathcal M}(\epsilon)$ be the union of all measurements of ${\mathcal M}$ within a distance $\epsilon$ centered on ${\mathcal M}$. Providing the limit exists, the Minkowski-Bouligand dimension of $\mathcal M$ is \cite{dubucetal}
\begin{equation}
\fdim^{\rm (MB)} = 2 - \lim_{\epsilon\to 0}\frac{\ln {\mathcal A}({\mathcal M}(\epsilon))}{\ln\epsilon}\,,
\label{fdim}
\end{equation}
where $\mathcal A$ is the area of ${\mathcal M}(\epsilon)$, i.e., the area traced out by a small circle with radius $\epsilon$ following the measurements ${\mathcal O}(t)$ from $0$ to $T$. It can be shown that Eq.\ (\ref{fdim}) can be rewritten as the box-counting dimension:
\begin{equation}
\fdim^{\rm (BC)} = \lim_{\epsilon\to 0}\frac{\ln {\mathcal N}(\epsilon)}{\ln(1/\epsilon)}\,,
\label{boxcounting}
\end{equation}
where $\mathcal N$ is the (minimum) number of disjoint squares of side $\epsilon$ that are necessary to cover ${\mathcal M}(\epsilon)$. The fractal dimension can be calculated from Eq.\ (\ref{fdim}) or Eq.\ (\ref{boxcounting}) by computing the slope of ${\mathcal A}({\mathcal M})$ or ${\mathcal N}(\epsilon)$ vs.\ $\epsilon$ in double logarithmic scale (log-log plot). For all purposes of this analysis, the Minkowski-Bouligand dimension and the box-counting dimension can be considered equivalent. Throughout the remainder of the paper we will make use of the definition in Eq.\ (\ref{fdim}), which will be referred as $\fdim$.

It was mentioned above that a set of discretely-sampled time-dependent measurements is not strictly a fractal. Therefore, the numerical evaluation of $\fdim$ with either method requires some care. A set of $N$ physical measurements of ${\mathcal M}$ is endowed with two scales that break the scaling invariance hypothesis at the basis of the fractal dimension derivation: A minimum scale defined by the inverse of the sampling frequency, $\Delta t=1/f_s$, and a maximum scale defined by the measurement time $T=N\Delta t$. This implies that the relation between $\ln A({\mathcal M})$ and $\ln\epsilon$ for physical sets is not linear, nor it can be calculated at scales smaller than $\Delta t$ or larger than $T$. In these cases, $\fdim$ is generally evaluated by computing the argument of the limit in Eq.\ (\ref{fdim}) at different scales $\epsilon_k=k\Delta t$, where $k=1,2,\dots M\le N$, and then performing a linear fit to evaluate the function in the limit of infinitely-sampled measurements, $f_s\to\infty$:
\begin{equation}
D_{F}(M) = 2 - {\bm{\mathcal S}}_M(\{\mathcal{P}_1,\dots \mathcal{P}_M\})\,,\label{dimension}
\end{equation}
where ${\bm{\mathcal S}}_M$ denotes the slope of the linear fit computed on the points ${\mathcal P}_k=(\ln\epsilon_k,\ln{\mathcal A}_k(\epsilon_k))$  corresponding to the $M$ possible scales in the data. 

An additional source of approximation in the calculation of $\fdim$ follows from the numerical scheme used to compute the numerator of Eq.\ (\ref{fdim}). In our analysis we employ two different methods to evaluate ${\mathcal A}({\mathcal M})$, the \ac{VAR} method \cite{Tricot1988EvaluationDL,dubucetal, VAR} and the ANAM method \cite{ANAM}. We follow Ref.\ \cite{Bigerelle2000b} for their implementation.

\begin{itemize}
\item{\bf VAR method.} The \ac{VAR} method \cite{Tricot1988EvaluationDL} is one of the most efficient techniques to evaluate the fractal dimension \cite{dubucetal, VAR}. The \ac{VAR} estimator is defined by taking the average of the function
\begin{equation}
{\mathcal F}[{\mathcal O}(t),\epsilon] = \Bigl| \hbox{max}\left\{t'\in[t-\epsilon,t+\epsilon],{\mathcal O}(t')\right\} -  \hbox{min}\left\{t'\in[t-\epsilon,t+\epsilon],{\mathcal O}(t')\right\} \Bigr|\,,\label{VARF}
\end{equation}
on the data set, i.e., 
\begin{equation}
{\mathcal A}^{\hbox{\bf\tiny{VAR}}}(\epsilon) = \frac{1}{T}\int_{0}^{T} dt\, {\mathcal F}[{\mathcal O}(t),\epsilon]\,.\label{VAR}
\end{equation}
For a discretly-sampled set of $N$ measurements at times $t_j=j\Delta t$, where $j=0,1,\dots N-1$, we evaluate ${\mathcal A}^{\hbox{\bf\tiny{VAR}}}(\epsilon)$ at the scales $\epsilon_k$. The discretized version of Eq.\ (\ref{VARF}) is a $d$-dimensional vector with components
\begin{equation}
{\mathcal F}_{k,l} = \Bigl| \hbox{max}[{\mathcal O}_{l-k}\dots{\mathcal O}_{l+k}] - \hbox{min}[{\mathcal O}_{l-k}\dots{\mathcal O}_{l+k}] \Bigr|\,,\label{VARFkl}
\end{equation}
where $l=[k,k+1,\dots N-k+1)$, $\mathcal{O}_j$ denotes the measurement at time $t_j$ and $M=N/2-1$. A simple numpy implementation returning the vector $\boldsymbol{\mathcal A}^{\hbox{\bf\tiny{VAR}}}$ with components ${\mathcal A}^{\hbox{\bf\tiny{VAR}}}_k$ is 
\begin{center}
\begin{minipage}{14cm}
\begin{python}
import numpy as np
def VAR(data):
   N = len(data)
   F_k = [[np.abs(np.max(data[l-k:l+k+1]) - np.min(data[l-k:l+k+1])) \
      for l in np.arange(l,N-l)] for k in np.arange(1,N//2)]
   A_var = [np.mean(np.asarray(F_k[j])) for j in np.arange(len(F_k))]
   return A_var
\end{python} 
\end{minipage}
\end{center}
Typically, the slope of $\boldsymbol{\mathcal A}^{\hbox{\bf\tiny{VAR}}}$ is constant to a good approximation. Therefore, it is not necessary to compute all its components. The computation of the estimator can be sped up by limiting the calculation of the second loop to {\tt N\,{\color{magenta}//}(2*decimate)}, where {\tt decimate} is a positive integer.

\item{\bf ANAM method.} The ANAM estimator \cite{ANAM} is defined by taking the average of the function
\begin{equation}
{\mathcal G}[{\mathcal O}(t),\epsilon] = \left[\frac{1}{\epsilon^2}\int_0^\epsilon dt_1 \int_0^\epsilon dt_2\, \bigl| {\mathcal O}(t+t_1)- {\mathcal O}(t-t_2)\bigr|^\alpha\right]^{1/\alpha}\label{ANAMG}
\end{equation}
on the data set, where $\alpha\ge 1$ is an arbitrary parameter, i.e.,
\begin{equation}
{\mathcal A}^{\hbox{\bf\tiny{ANAM}}}(\epsilon) = \frac{1}{T}\int_{0}^{T}\,dt\, {\mathcal G}[{\mathcal O}(t),\epsilon]\,.\label{ANAM}
\end{equation}
Equation (\ref{ANAM}) can be calculated as a discretized vector at the scales $\epsilon_k$:
\begin{equation}
{\mathcal A}^{\hbox{\bf\tiny{ANAM}}}_k = \frac{1}{N-2k}\sum_{j=k}^{N-k-1}\left[\frac{1}{(k+1)^2}\sum_{i,l=0}^k\bigl|\mathcal{O}_{j+i}-\mathcal{O}_{j-l}\bigr|^\alpha\right]^{1/\alpha}\,,\label{ANAM_num}
\end{equation}
A simple numpy implementation returning the $\boldsymbol{\mathcal A}^{\hbox{\bf\tiny{ANAM}}}$ vector with components ${\mathcal A}^{\hbox{\bf\tiny{ANAM}}}_k$ is 
\begin{center}
\begin{minipage}{14cm}
\begin{python}
import numpy as np
def ANAM(data):
   N = len(data)
   df = np.asarray([np.power(np.abs(data - data[i]),args.alpha) \
      for i in np.arange(N)]).reshape(N,N)
   A_anam = [(k+1)**(-2/alpha)/(N-2*k)*np.sum([np.power(np.sum([df[j+i,j-l] \
      for i in np.arange(k+1) for l in np.arange(k+1)]),1/alpha) \
      for j in np.arange(k,N-k)]) for k in np.arange(1,N//2)]
   return A_anam 
\end{python} 
\end{minipage}
\end{center}
Similarly to the VAR estimator, the computation of the ANAM estimator can be sped up by limiting the calculation of the last loop to {\tt N\,{\color{magenta}//}(2*decimate)}.
Moreover, it can be shown that the theoretical fractal dimension in Eq.\ (\ref{ANAM}) does not depend on the value of $\alpha$ \cite{ANAM}. In the following we will set $\alpha=1$ to minimize the
numerical complexity of Eq.\ (\ref{ANAM_num}) and speed up the numerical calculation.

\end{itemize}

Using either estimator and Eq.\ (\ref{dimension}), a numerical estimate of the fractal dimension is obtained by computing the slope of the curve traced by the points $(\ln k,\ln{\mathcal A}_k)$. Although the curve is expected to be linear, this is only approximately true because of the finite sampling rate and length of the data, the numerical approximations used to compute ${\mathcal A}_k$, and machine computational errors. Therefore, the fractal dimension is estimated by extracting  the slope of the curve with a linear fit.  

\section{Algorithm Performance Tests}\label{tests}

\begin{figure}[ht]
\begin{center}
\includegraphics[width=0.45\textwidth]{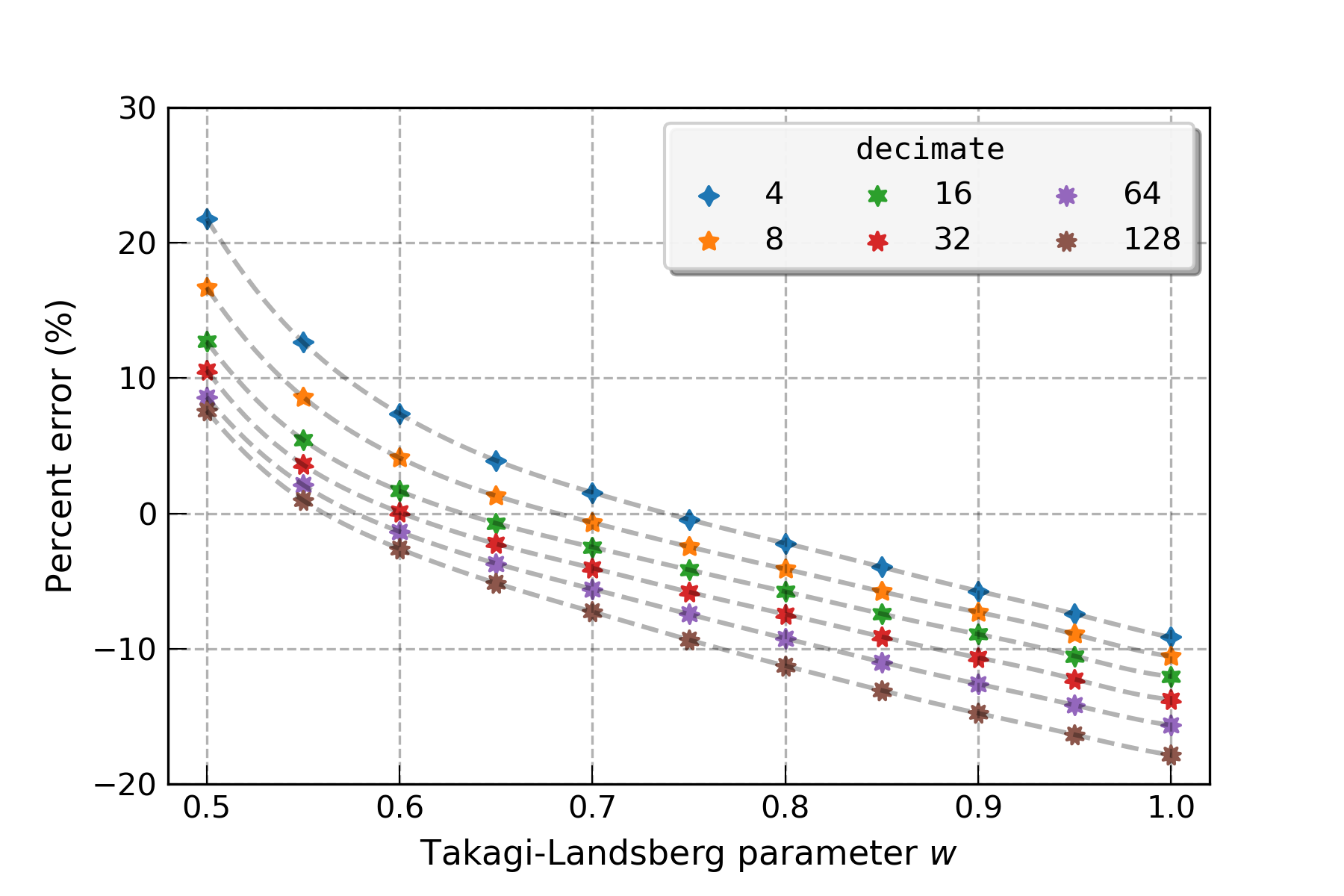}~
\includegraphics[width=0.45\textwidth]{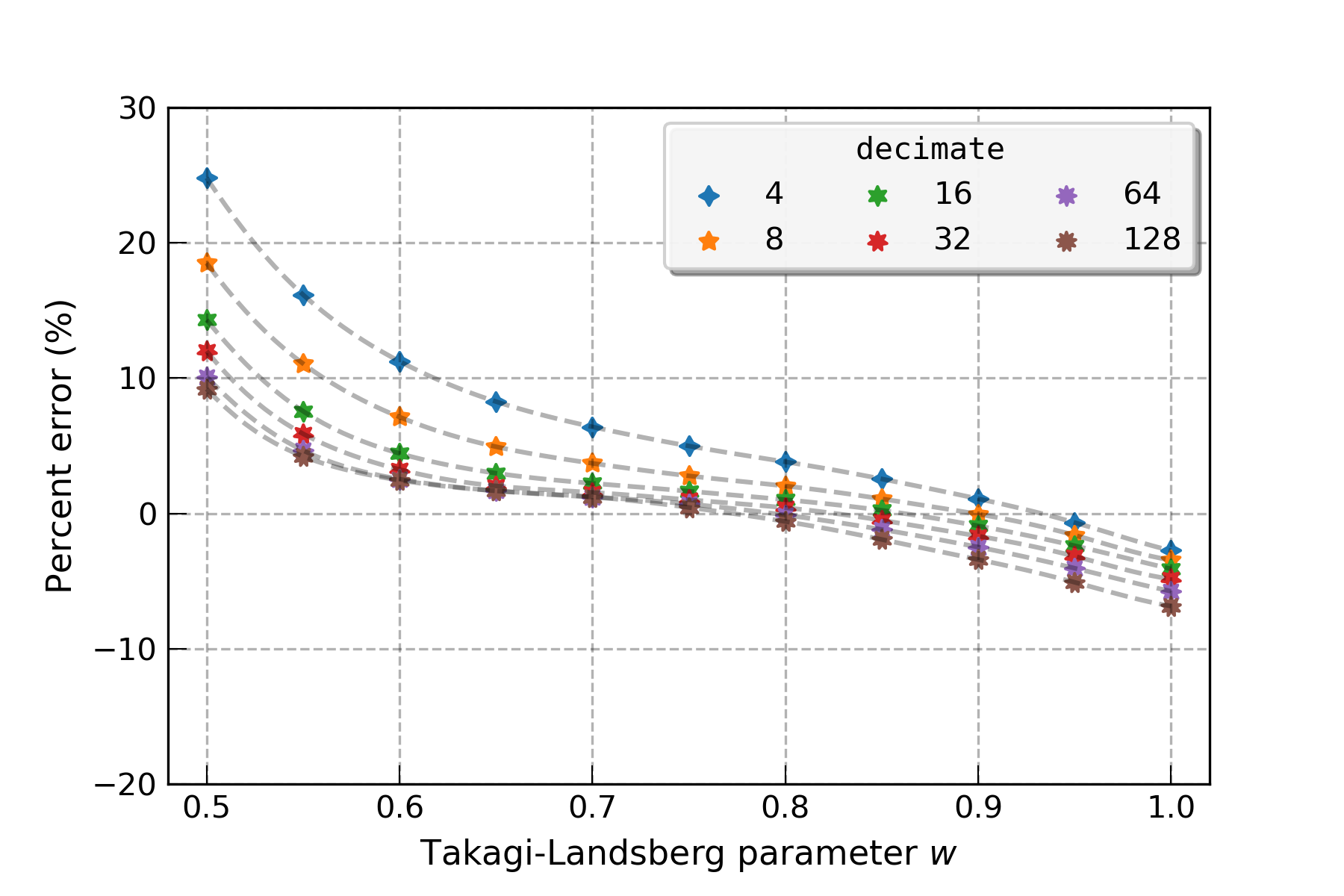}
\caption{Percent errors for the fractal dimension of the \ac{TL} function for different values of the parameter $w$ and {\tt decimate}. Left: VAR estimator. Right: ANAM estimator. Fits are done with the Gaussian Process Regressor implementation of sklearn \cite{scikit-learn} with kernel {\tt RBF(length\_scale=10,length\_scale\_bounds=[1e-01, 10.0])}.}
\label{TK}
\end{center}
\end{figure}

\begin{figure}[ht]
\begin{center}
\includegraphics[width=0.45\textwidth]{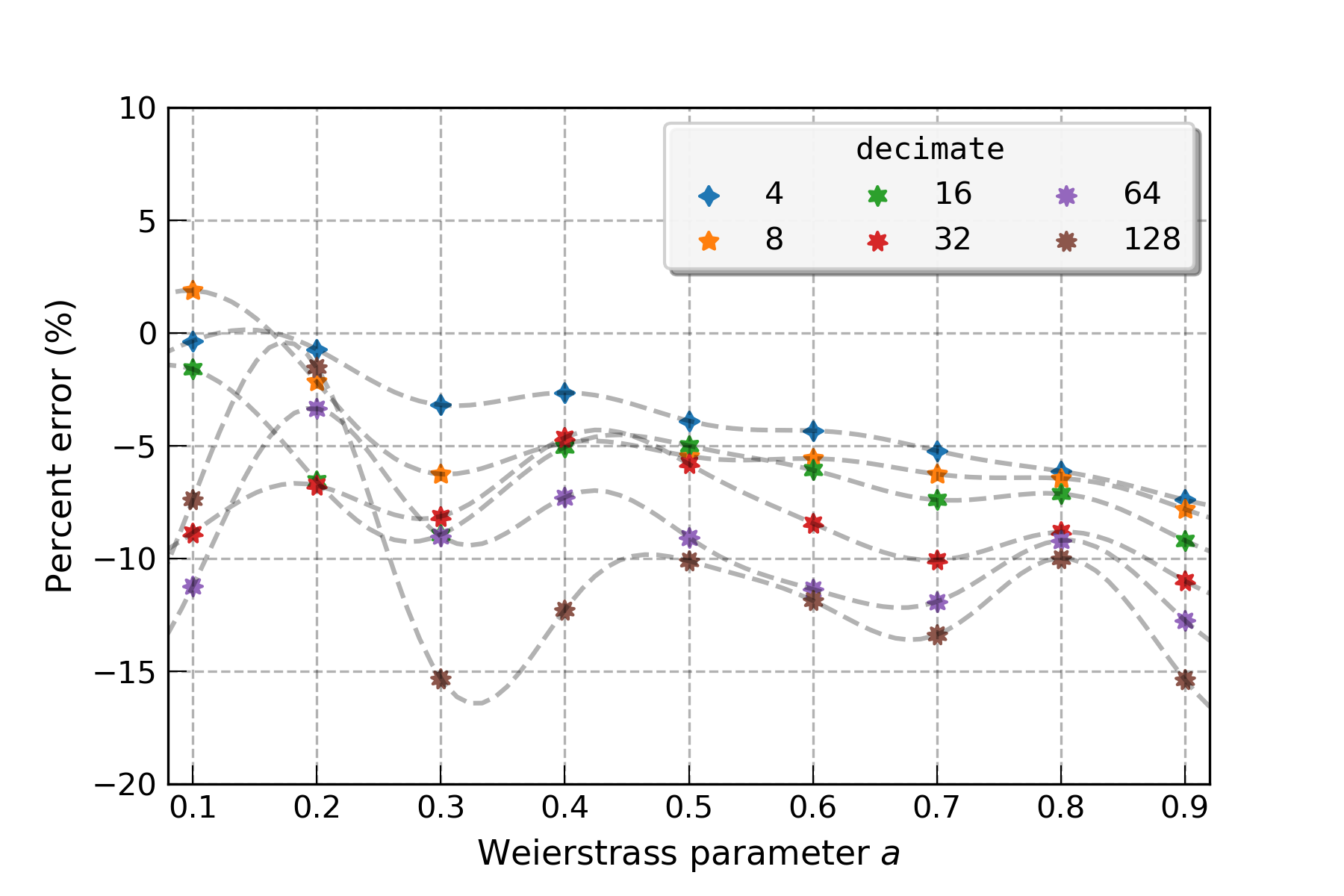}~
\includegraphics[width=0.45\textwidth]{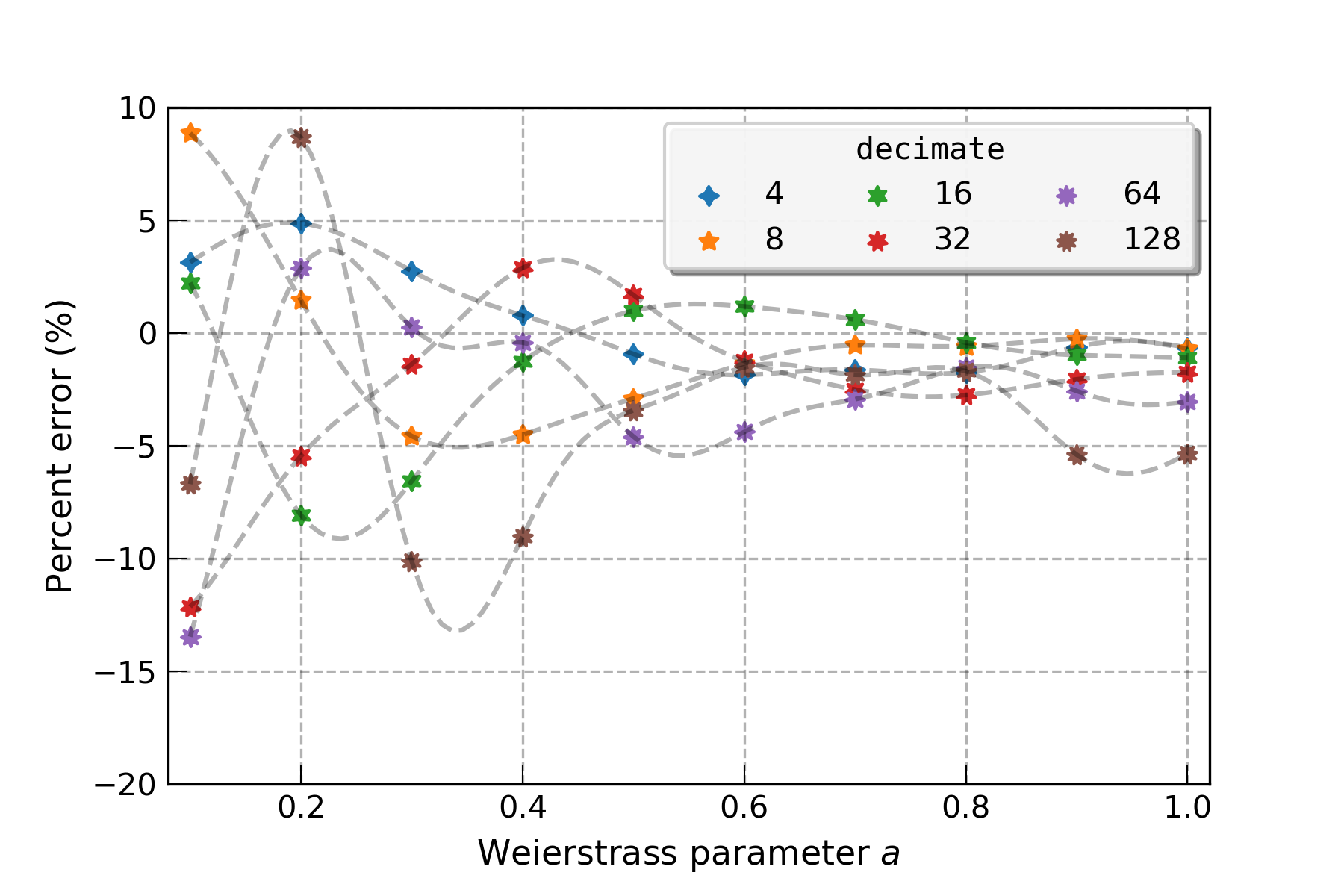}
\caption{Percent errors for the fractal dimension of the \W function for different values of the parameter $a$ and {\tt decimate}. The parameter $b$ is fixed as {\tt b = int((1.+1.5*np.pi)/a+1.)}. Left: VAR estimator. Right: ANAM estimator. Fits are done with the Gaussian Process Regressor implementation of sklearn \cite{scikit-learn} with kernel {\tt RBF(length\_scale=10,length\_scale\_bounds=[1e-01, 10.0])}.}
\label{Weierstrass}
\end{center}
\end{figure}

We test the performance of the estimators by evaluating $\fdim$ on a few real-valued data sets with known fractal dimension. The \ac{TL} function and the \W function are two well-known fractal sets whose fractal dimension can be computed analytically. The \ac{TL} function is defined as 
\begin{equation}
T_w(t) = \sum_{n=0}^\infty w^n s(2^nt)\,,
\end{equation}
where $|w|<1$ is a real parameter and $s(t)$ is the distance from $t$ to the nearest integer:
\begin{equation}
s(t)=\begin{aligned}
{\rm min}\\[-0.8em]
{\scriptscriptstyle n\in \mathbb{Z}\;}
\end{aligned}~|t-n|\,.
\end{equation}
The fractal dimension of the \ac{TL} function is
\begin{equation}
D_{TL}(w) = 2+\log_2(w)\,.
\end{equation}
The \W function is defined as
\begin{equation}
W_{ab}(t) = \sum_{k=1}^\infty a^{-k}\sin(b^kt)\,,
\end{equation}
where $1<a<2$ and $b>1$ are real parameters. The fractal dimension of the \W function is
\begin{equation}
D_{W}(a,b)=2-\log_b(a)\,.
\end{equation}
Figures \ref{TK} and \ref{Weierstrass} show the percent errors of the fractal dimension calculated with the VAR and ANAM methods for one second-long \ac{TL} and \W time series sampled at 4096 Hz,
different function parameters and different values of {\tt decimate}, respectively. Errors in the estimate of $\fdim$ are typically within a few percent with the largest values occurring at the
boundaries of the function convergence intervals ($w=1/2,1$ for the \ac{TL} function and $a=0,1$ for the \W function). These tests also show that the ANAM estimator is generally slightly more
accurate than the VAR estimator.  The price one needs to pay for this better accuracy is in a lower computational speed of the ANAM estimator compared to the VAR estimator. The simple python
implementation of the ANAM method shown in the previous section can be slower by as much as two orders of magnitude than the VAR implementation, depending on the value of {\tt decimate}. The
accuracy of the estimate depends on the value of {\tt decimate}, the region in the function's parameter space and the estimator. For example, percent errors with {\tt decimate}=4 tend to be smaller
than the corresponding errors with {\tt decimate}=128 for the \W function for most of its parameter space, irrespective of the estimator used. Since in our study we are concerned with time
variations of the fractal dimension rather than its absolute value, the main factor determining the choice of {\tt decimate} is computational speed.   

\begin{figure}[ht]
\begin{center}
\includegraphics[width=0.45\textwidth]{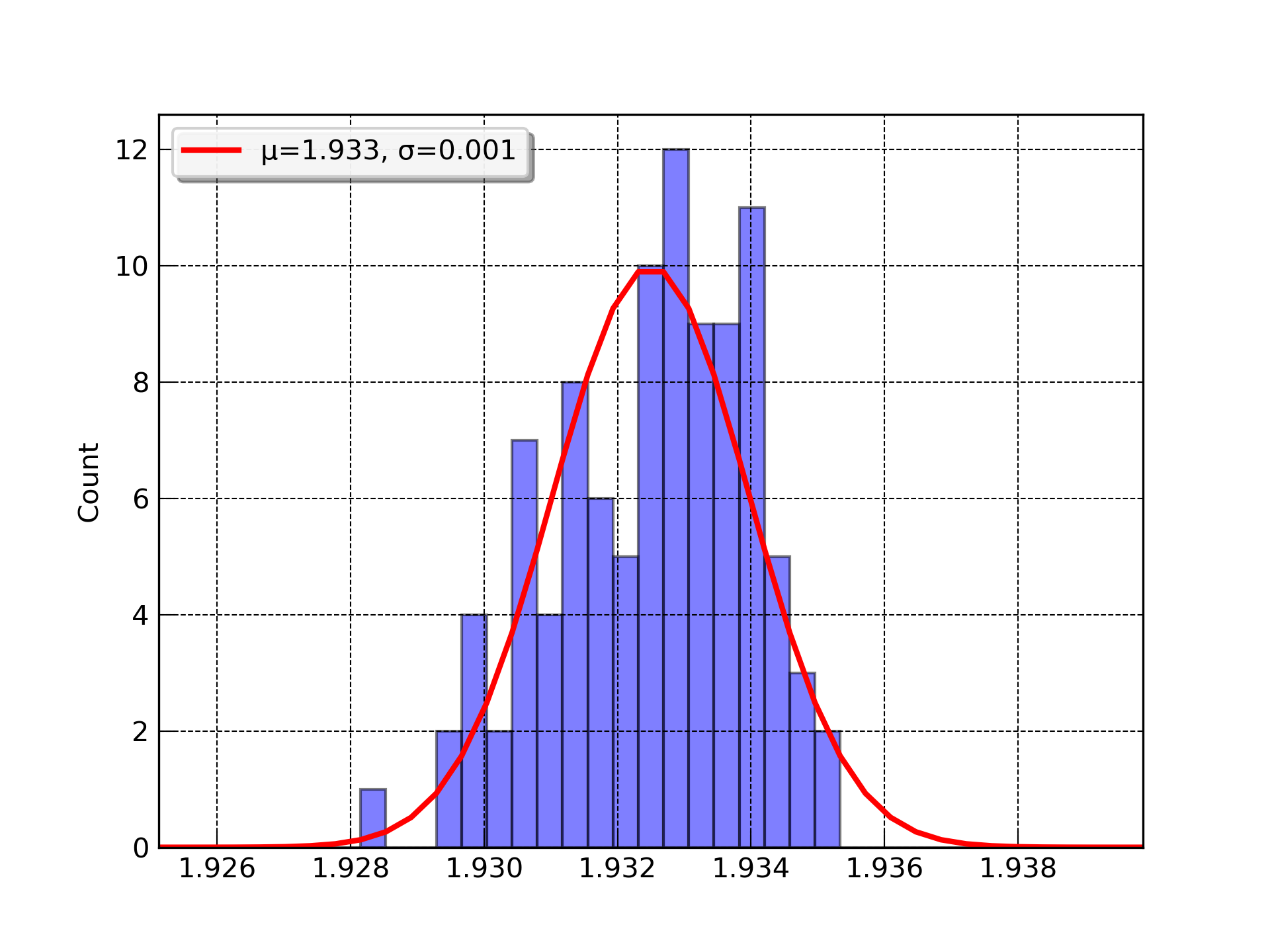}~
\includegraphics[width=0.45\textwidth]{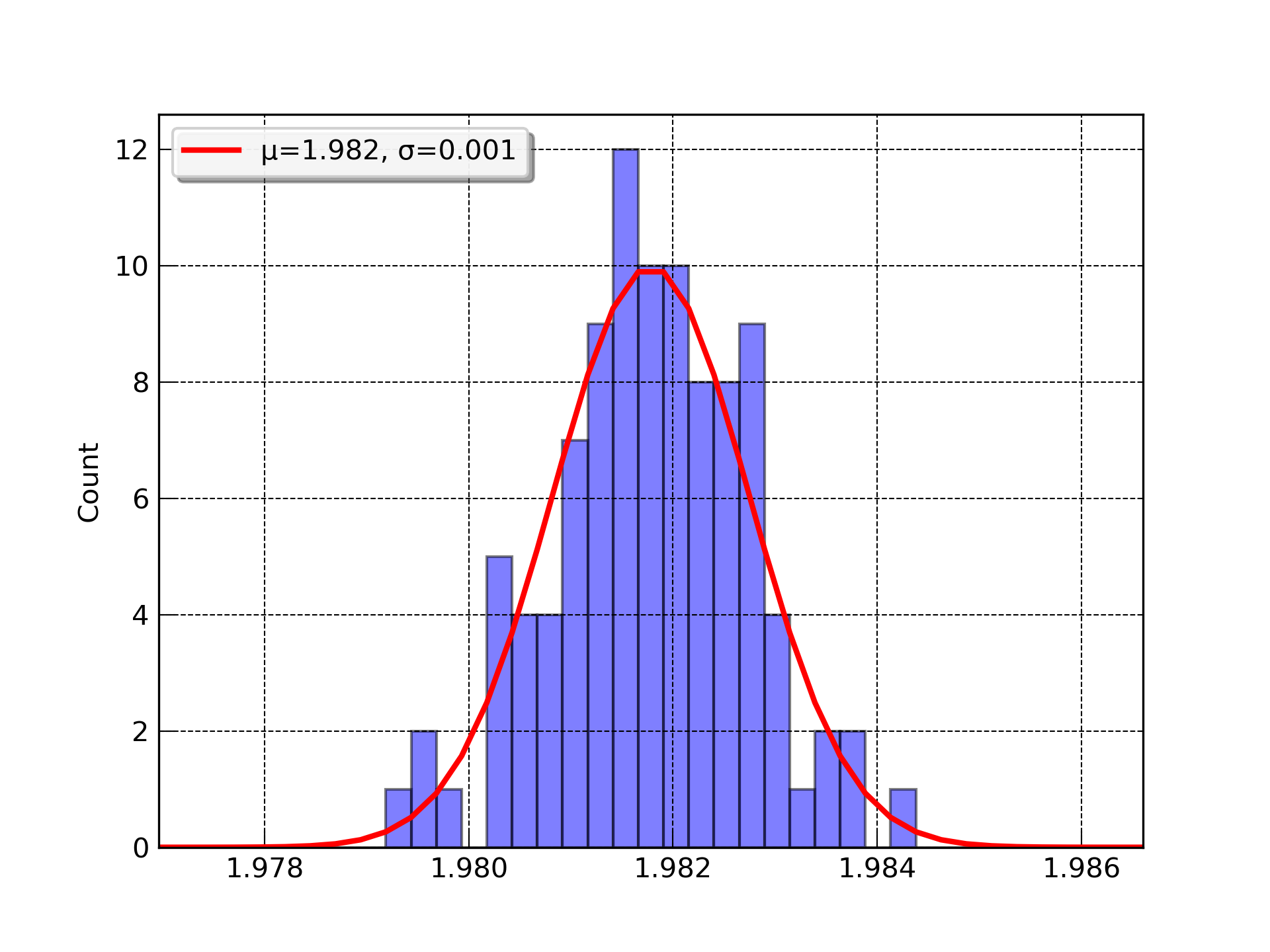}
\caption{Distribution of the fractal dimension for one hundred, one second-long white noise series sampled at 4096 Hz. The parameter {\tt decimate} is 16. Left: VAR estimator. Right: ANAM estimator. Fits are Gaussian curves with mean $\mu$ and standard deviation $\sigma$.}
\label{Whitenoise}
\end{center}
\end{figure}

\begin{figure}[ht]
\begin{center}
\includegraphics[width=0.45\textwidth]{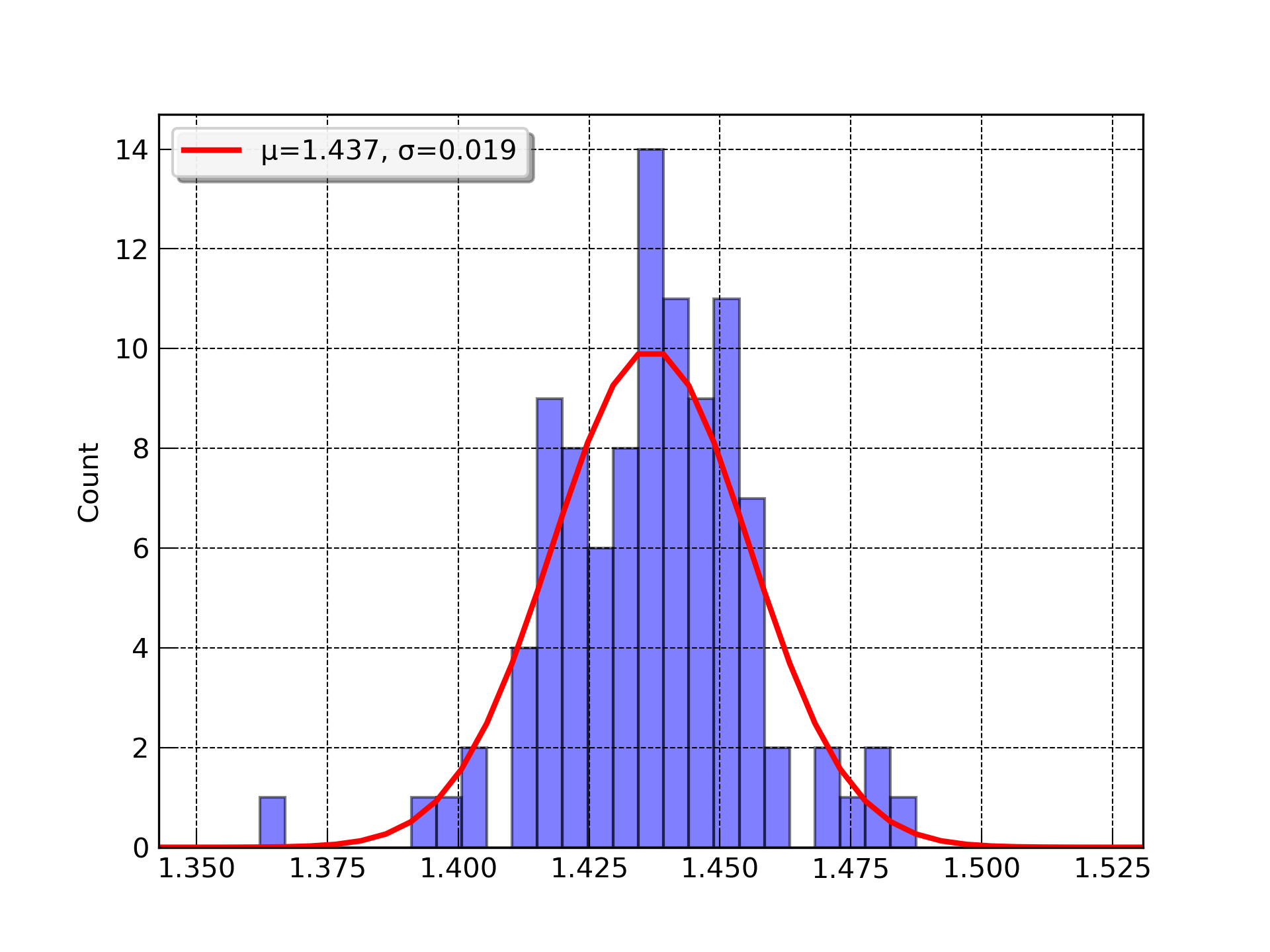}~
\includegraphics[width=0.45\textwidth]{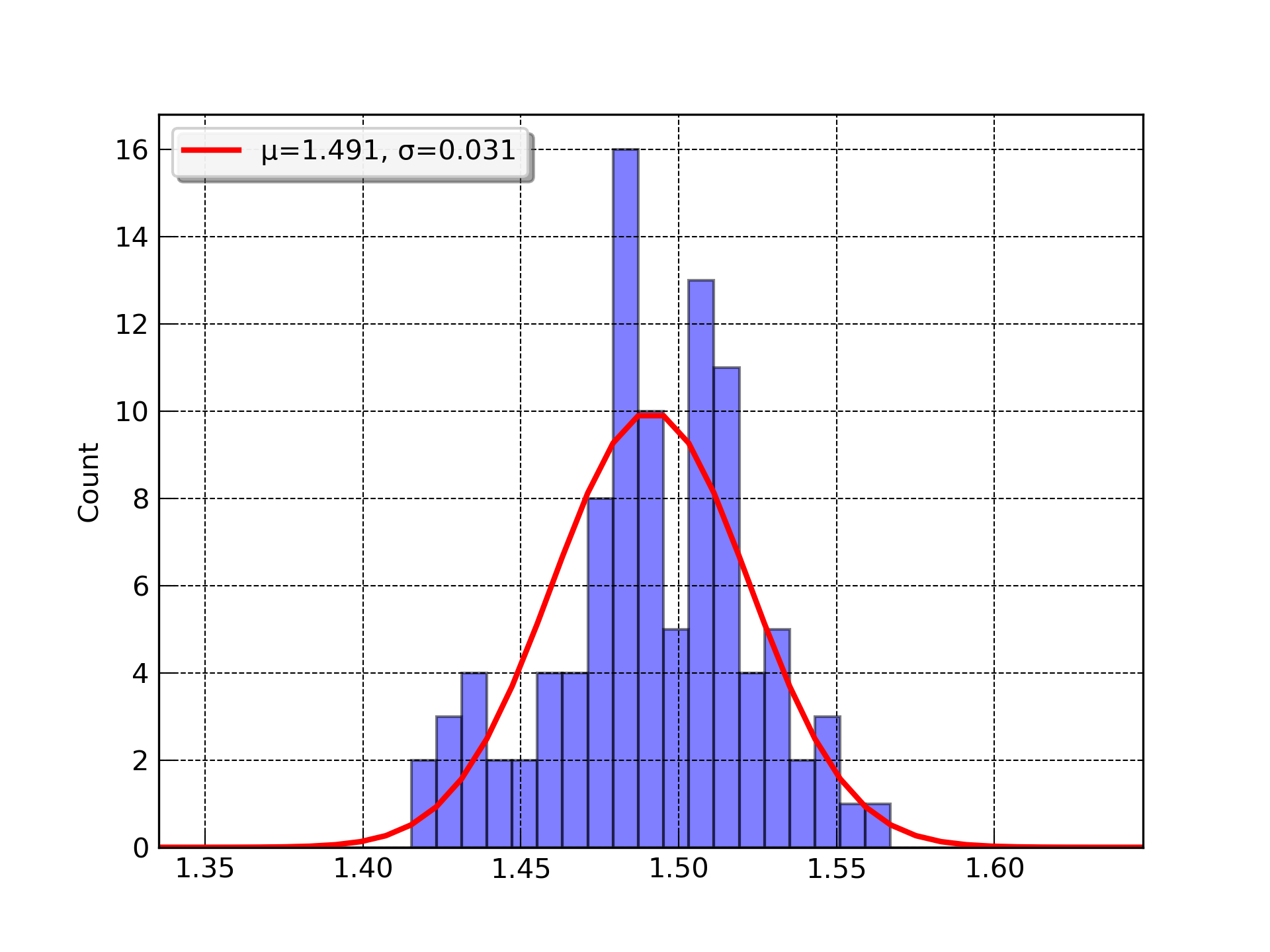}
\caption{Distribution of the fractal dimension for one hundred, one second-long Brownian noise series sampled at 4096 Hz. The parameter {\tt decimate} is 16.  The speed of the Brownian process is 2. Left: VAR estimator. Right: ANAM estimator. Fits are Gaussian curves with mean $\mu$ and standard deviation $\sigma$.}
\label{Brownnoise}
\end{center}
\end{figure}

Next we test the performance of the algorithms on time series which better resemble the noise of \ac{GW} detectors: white noise and Brownian noise \cite{10.5555/3002824}. Figure \ref{Whitenoise}
shows the distributions of the fractal dimension for a set of one hundred, one second-long white noise series sampled at 4096 Hz calculated with the VAR and ANAM methods and {\tt decimate}=16. The
percent error from the theoretical value $\fdim=2$ is about -3\% for the VAR estimator and -0.9\% for the ANAM estimator. Figure \ref{Brownnoise} shows the corresponding results for Brownian noise.
In this case, the percent errors from the theoretical value $\fdim=1.5$ are about -4\% and -0.6\% for the VAR and ANAM estimators, respectively. Similarly to the \ac{TL} and \W functions, the ANAM
estimator seems to be slightly more accurate than the VAR estimator. 

\section{Application to LIGO data}

To evaluate the performance of the algorithm on interferometer data and show its effectiveness in identifying noise transients and monitoring the stationarity of the detector we apply the VAR
method on two four-hour stretches of \ac{LIGO} data from the third LVK observing run \cite{gwosc}. In the first example, we choose a period of time where the \ac{L1} interferometer noise was
contaminated by a class of radio frequency beat note noise transients, colloquially called \emph{whistles} \cite{Davis_2021}. In the second example, we consider a different period of time where
data were corrupted by a different class of noise transients caused by stray scattered light in the interferometer optical system \cite{Davis_2021}. In our analysis, we focus on the strain channel
and one of the main auxiliary channels that witness the noise transients. We will show  that the fractal dimension method allows for the identification of the noise transients for both channels and
both glitch classes.

Whistle glitches are broadband noise transients with frequency ranging from less than $\sim 100$ Hz to several kHz with typical duration of the order of $\sim$ 1 second. Their time-frequency
representation shows the typical pattern of a frequency-decreasing arch followed by a frequency-increasing arch. Whistle glitches are caused by beat notes in the instrument Voltage Controlled
Oscillators and are witnessed by a number of auxiliary channels depending on their origin. For the stretches of data considered here, the main witness channel identified by \emph{hveto}
\cite{Smith_2011} is one of the fast sensing control auxiliary channels used to monitor the interferometer \ac{PRC}, {\tt L1:LSC-PRCL\_OUT\_DQ}. (For a list of \ac{LIGO} abbreviations and acronyms, and naming conventions for auxiliary channel and data quality flags naming conventions, see \cite{gwosc}.)

\begin{figure}[ht]
\begin{center}
\includegraphics[width=0.45\textwidth]{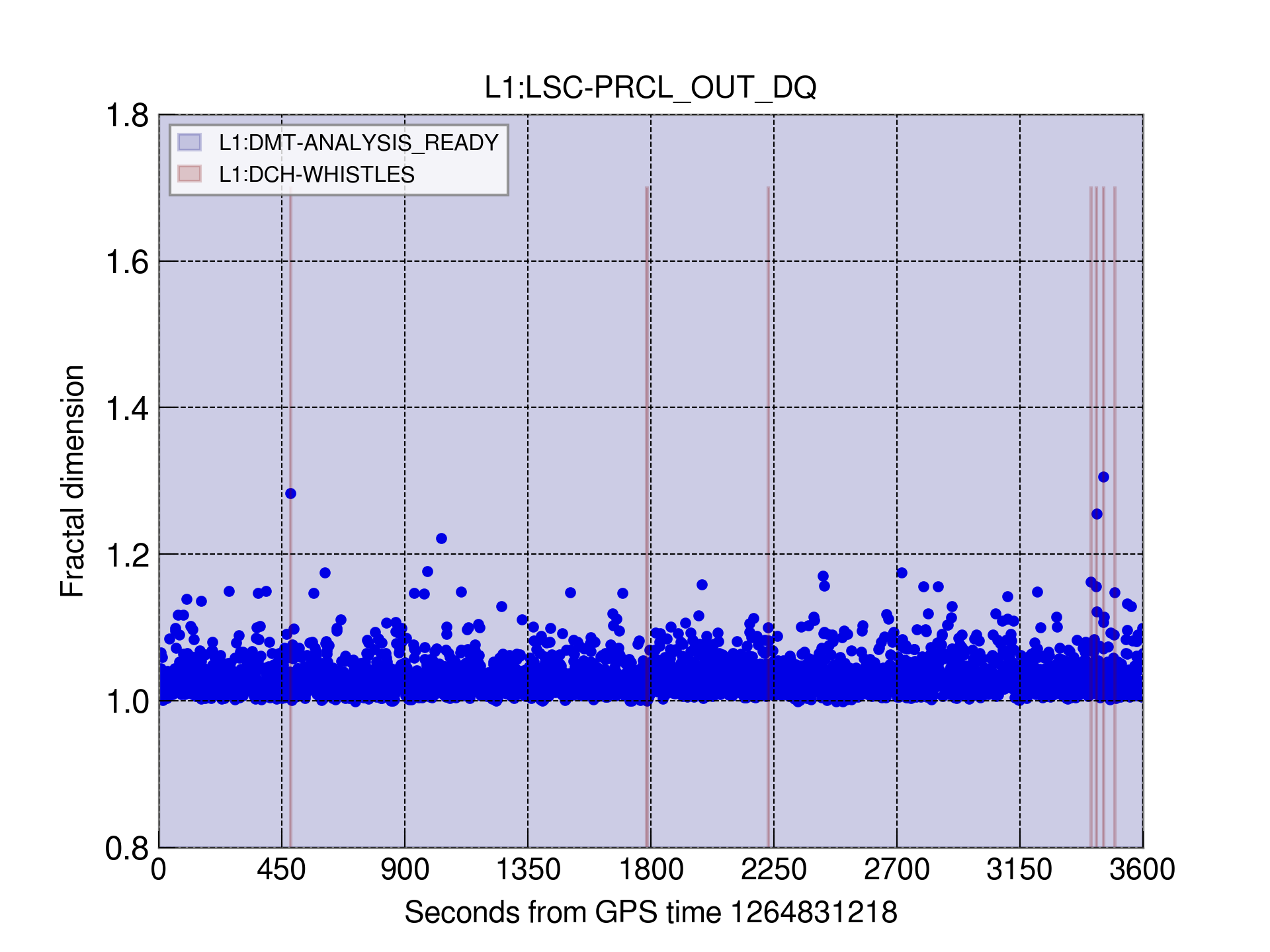}~
\includegraphics[width=0.45\textwidth]{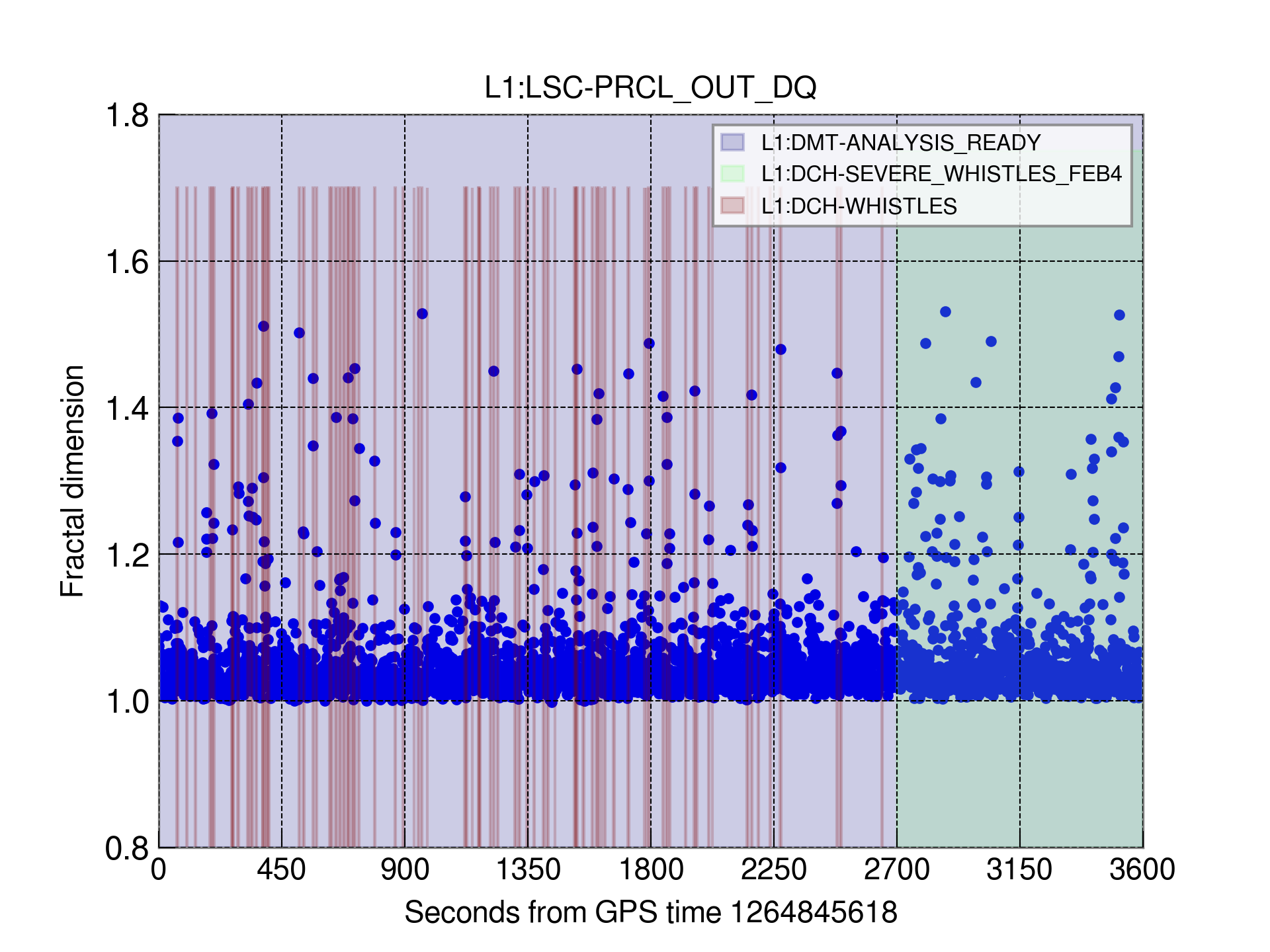}\\
\includegraphics[width=0.45\textwidth]{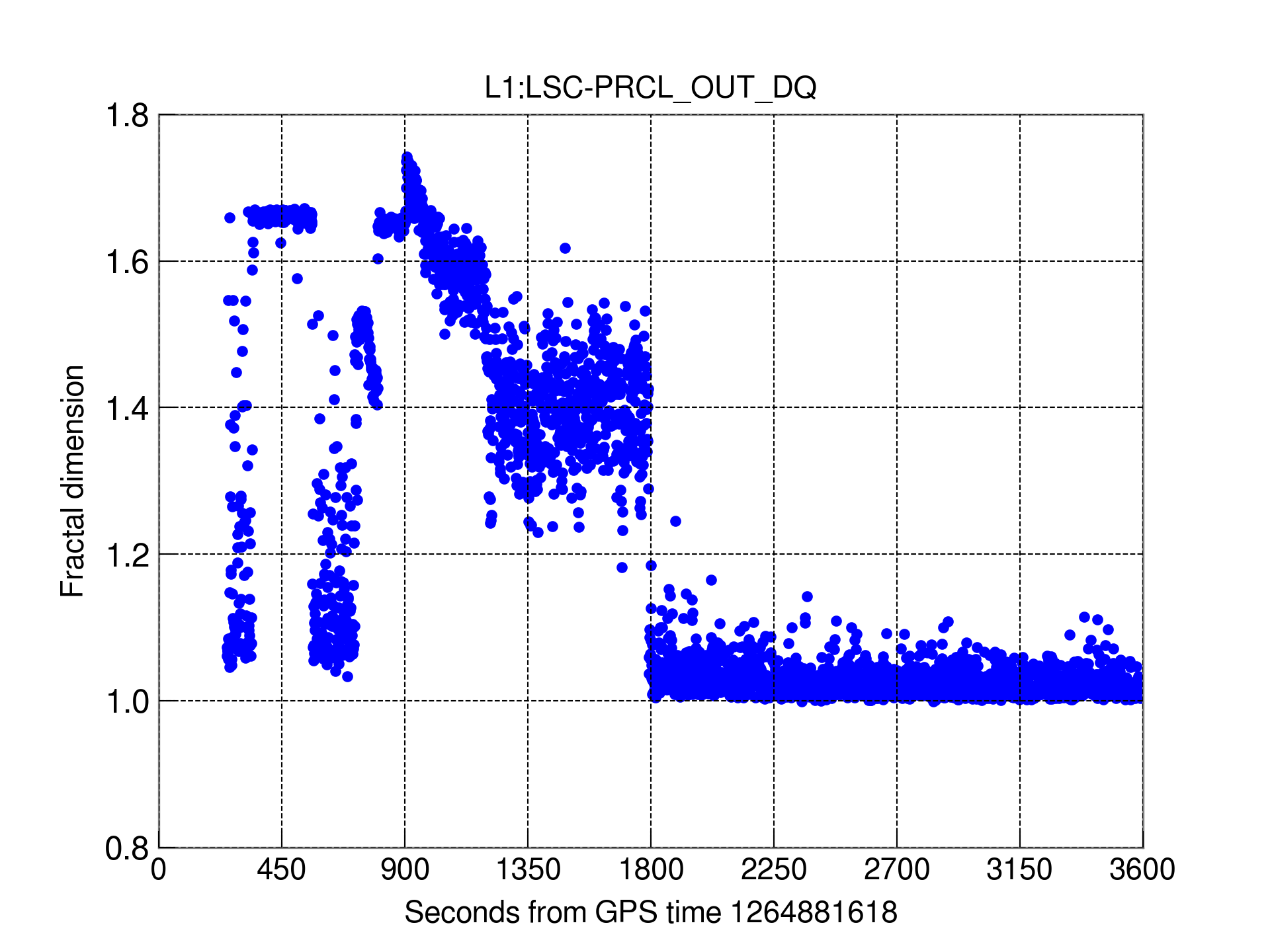}~
\includegraphics[width=0.45\textwidth]{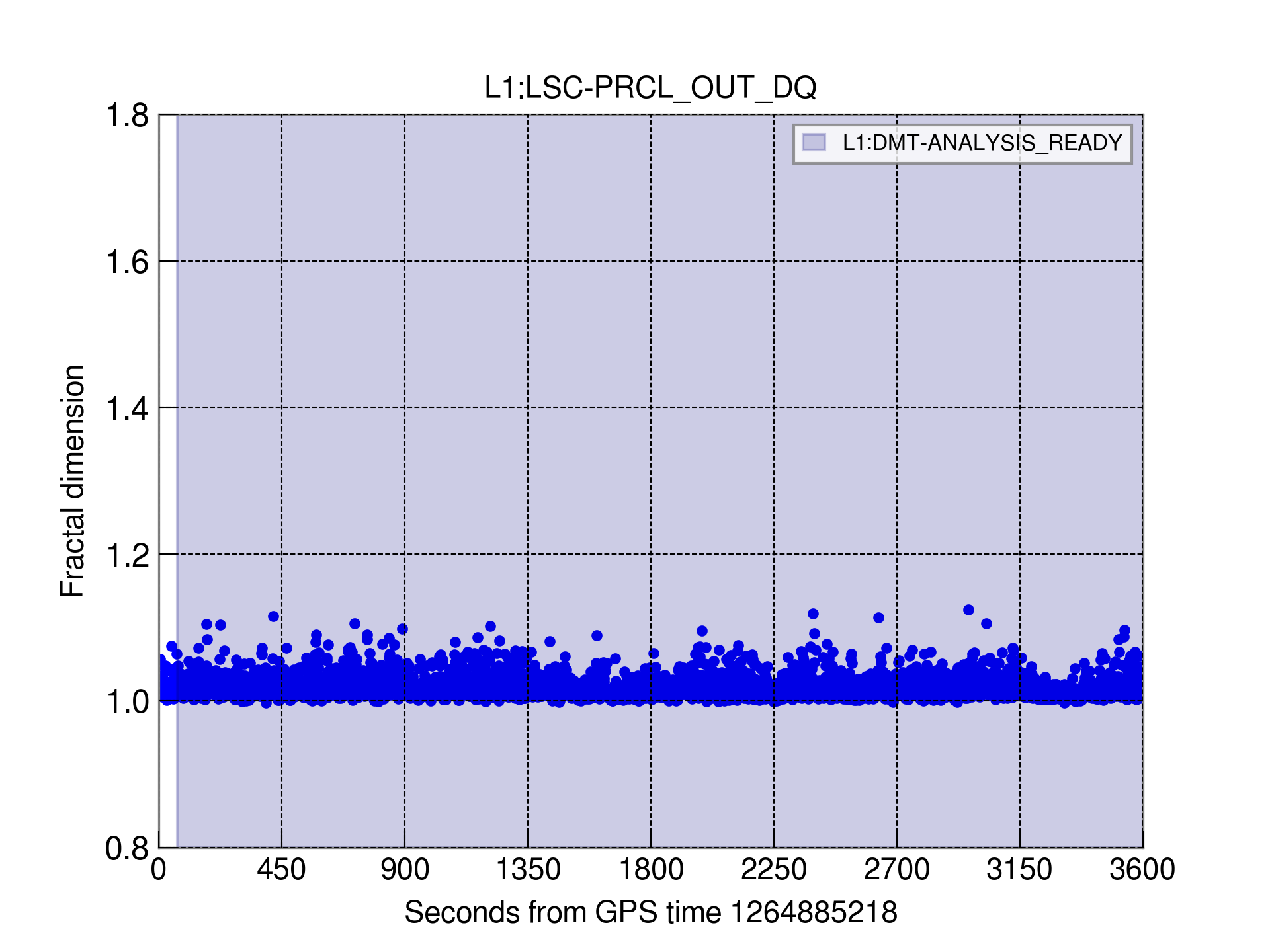}
\caption{Fractal dimension of four, one-hour periods of \ac{L1} data for the {\tt L1:LSC-PRCL\_OUT\_DQ} auxiliary channel (a witness of whistle glitches). The sampling rate of the channel is 16,384 Hz. The fractal dimension is computed with the VAR algorithm decimated at 64. Each point represents $\fdim$ for one second of data. The interferometer is in observing mode ({\tt L1:DMT-ANALYSIS\_READY}) during the periods corresponding to the top-left, top-right and bottom-right panels, and out of observing mode during the period corresponding to the bottom-left panel. Different color shades denote different data quality flags.}
\label{FD-L1_LSC-PRCL_OUT_DQ-var-dec_64-start_2020-02-04}
\end{center}
\end{figure}

\begin{figure}[ht]
\begin{center}
\includegraphics[width=0.45\textwidth]{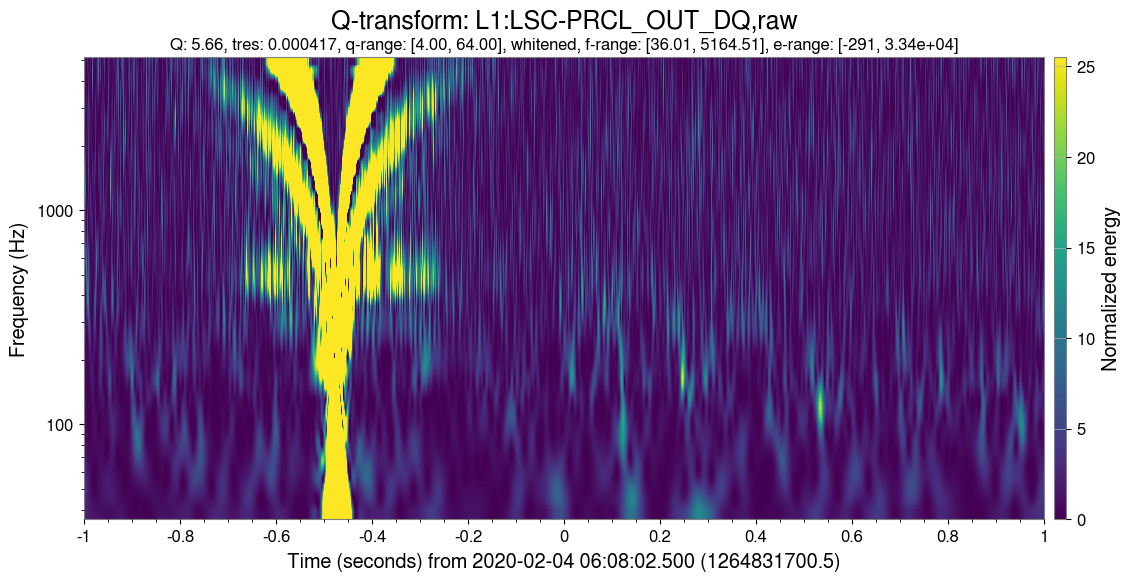}~
\includegraphics[width=0.45\textwidth]{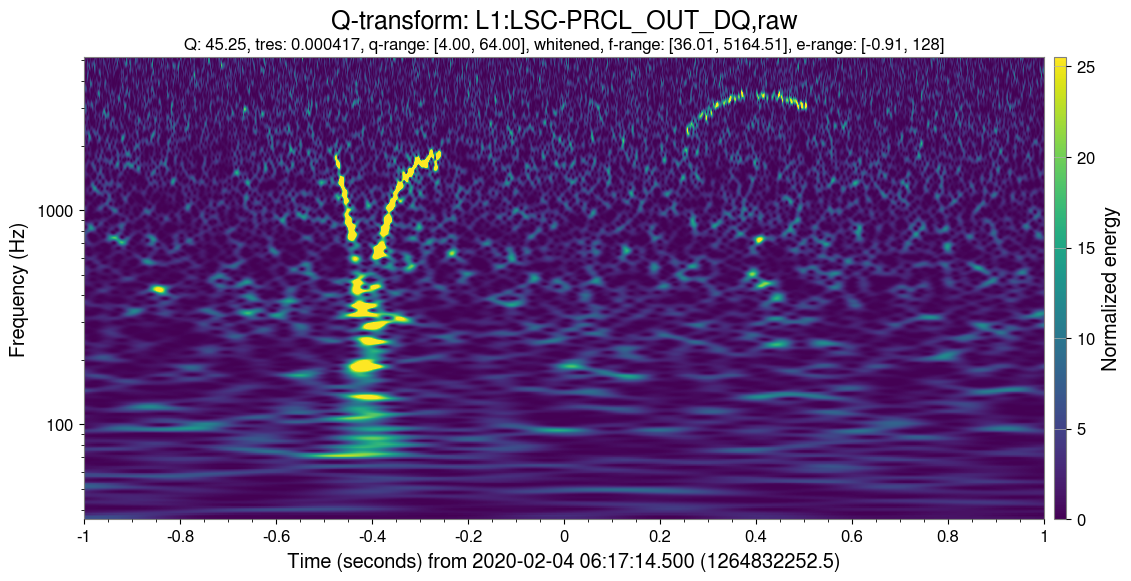}\\
\includegraphics[width=0.45\textwidth]{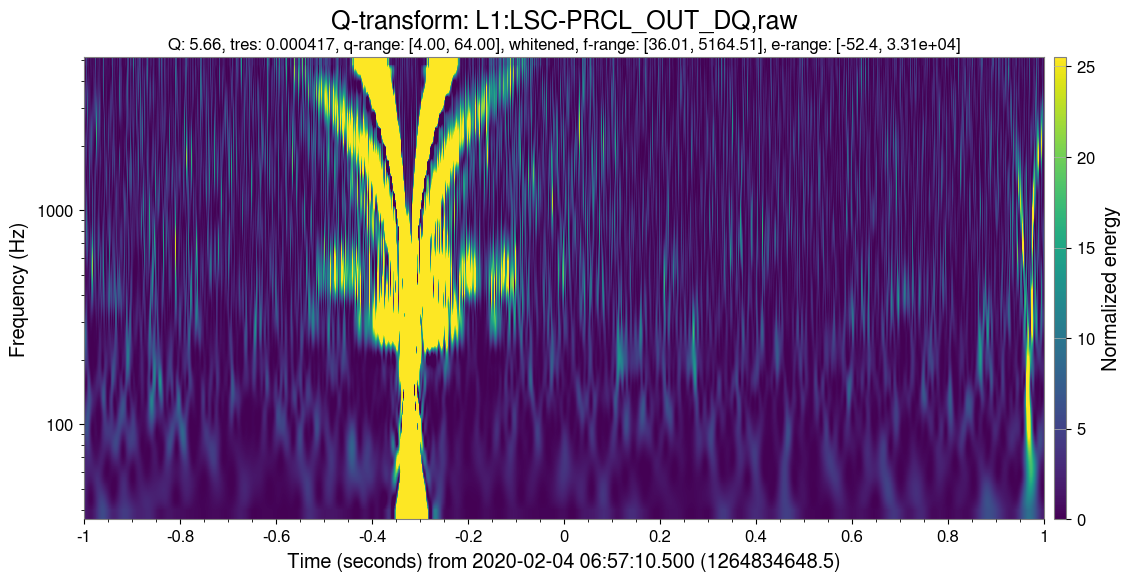}~
\includegraphics[width=0.45\textwidth]{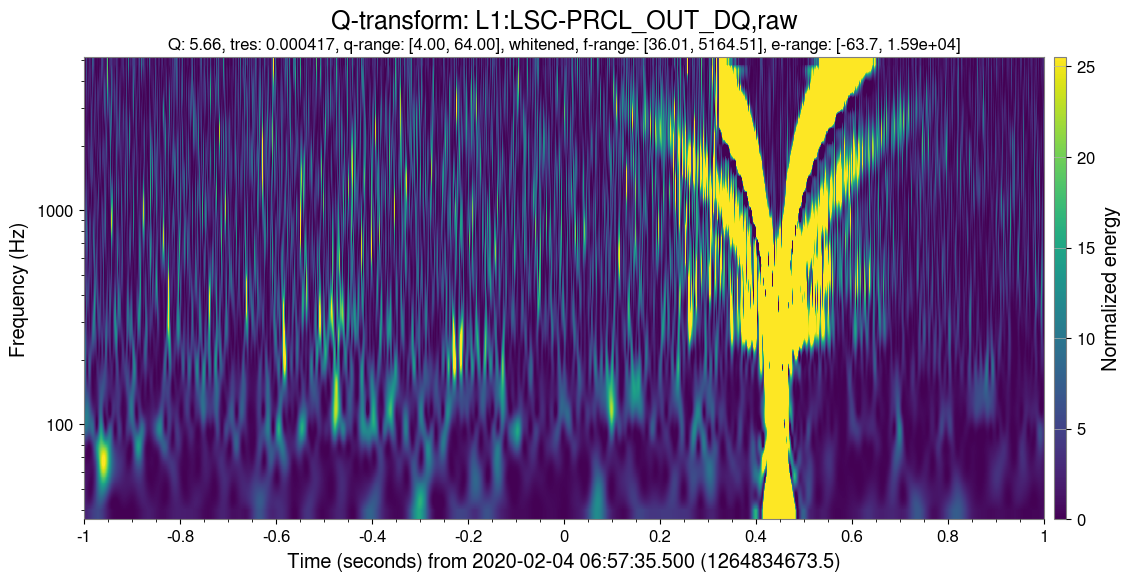}
\caption{Q-scans of the {\tt L1:LSC-PRCL\_OUT\_DQ} auxiliary channel centered on the four anomalous points with $\fdim>1.2$ in the top-left panel of Fig.~\ref{FD-L1_LSC-PRCL_OUT_DQ-var-dec_64-start_2020-02-04} showing the presence of whistle glitches. The top-left panel corresponds to the anomalous point at $t=482$ seconds from the beginning of the one-hour time period. The top-right panel corresponds to the anomalous point with value closer to the $\fdim=1.2$ threshold occurring at $t=1034$ seconds from the beginning of the one-hour time period.
The bottom panels correspond to the anomalous points occurring a few minutes before the end of the one-hour period.}
\label{QSCAN-L1_LSC-PRCL_OUT_DQ-var-dec_64-start_2020-02-04}
\end{center}
\end{figure}

\begin{figure}[ht]
\begin{center}
\includegraphics[width=0.45\textwidth]{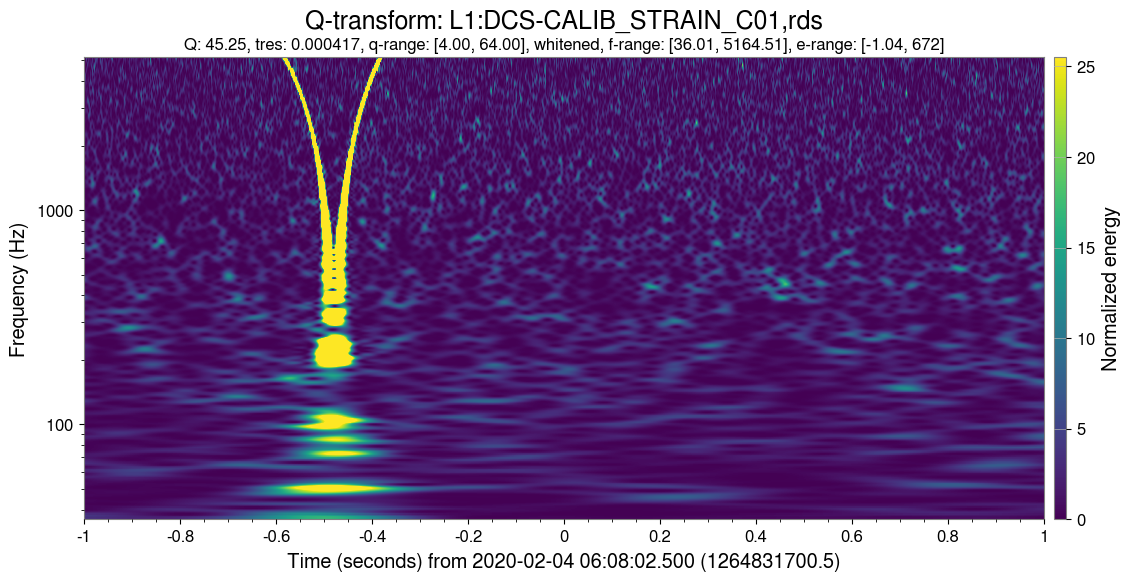}~
\includegraphics[width=0.45\textwidth]{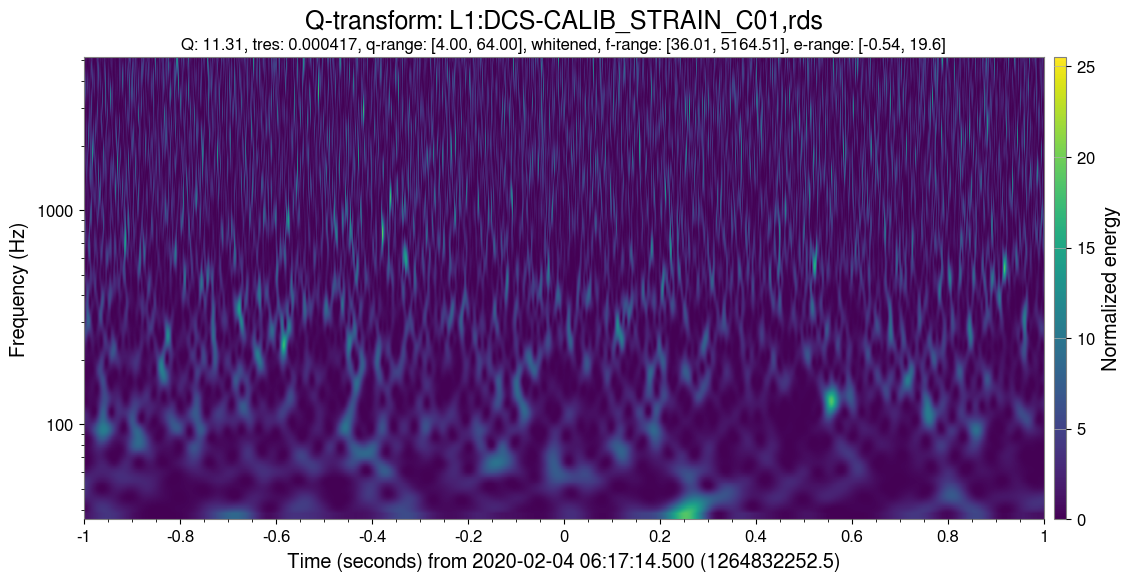}\\
\includegraphics[width=0.45\textwidth]{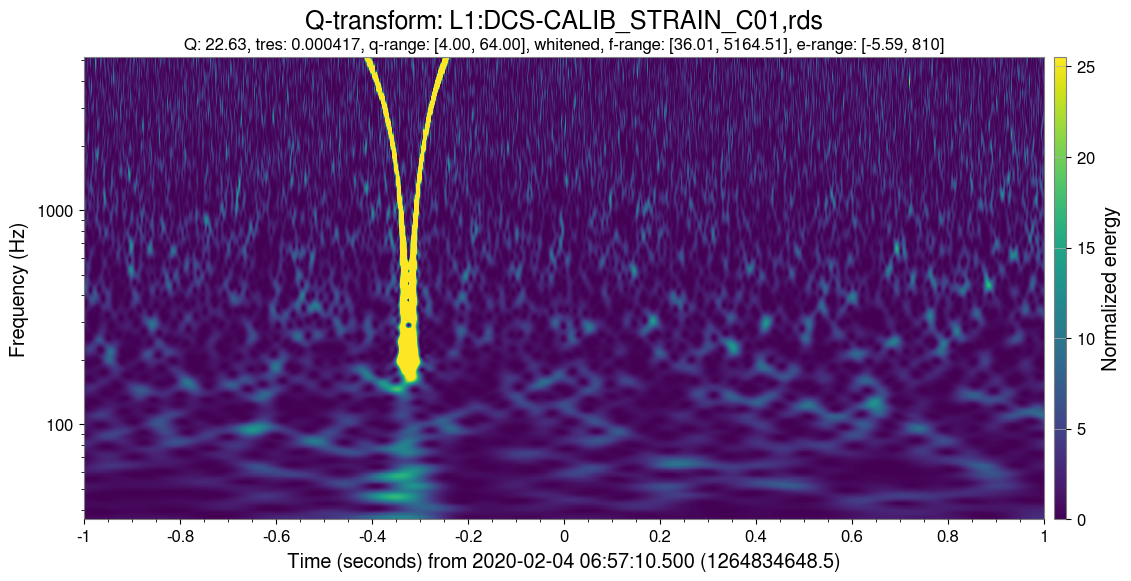}~
\includegraphics[width=0.45\textwidth]{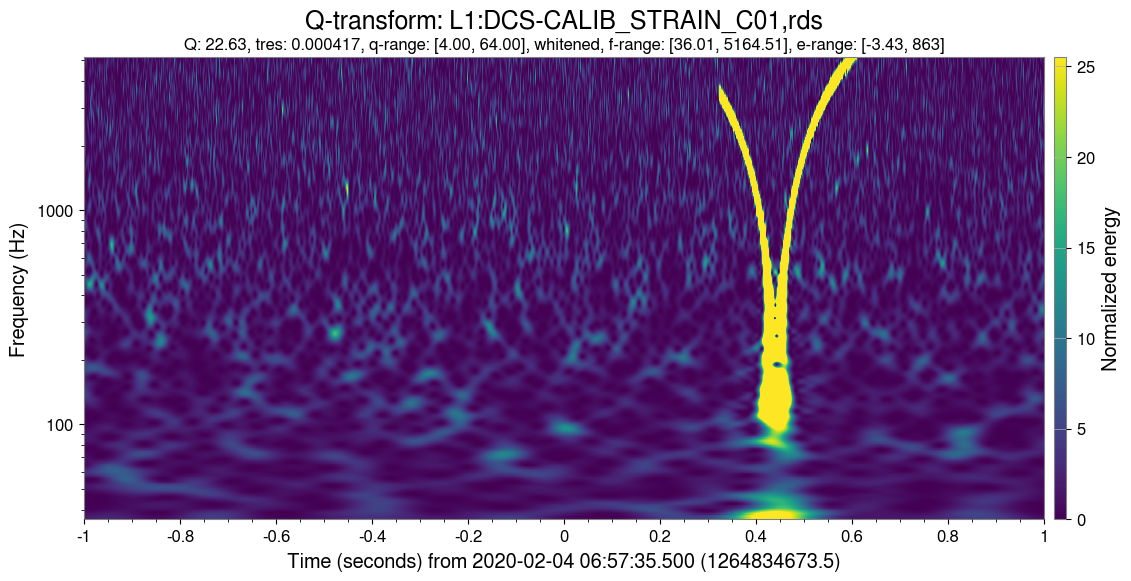}
\caption{Q-scans of the calibrated strain channel {\tt L1:DCS-CALIB\_STRAIN\_C01} for the four glitches of Fig.~\ref{QSCAN-L1_LSC-PRCL_OUT_DQ-var-dec_64-start_2020-02-04}. The whistle glitch in the top-right panel is not visible due to its lower \ac{SNR} although it is apparent in the Q-scan of the {\tt L1:LSC-PRCL\_OUT\_DQ} auxiliary channel.}
\label{QSCAN-DCS-CALIB_STRAIN_C01-var-dec_64-start_2020-02-04}
\end{center}
\end{figure}

Figure \ref{FD-L1_LSC-PRCL_OUT_DQ-var-dec_64-start_2020-02-04} shows the evolution of $\fdim$ over four hours of the {\tt L1:LSC-PRCL\_OUT\_DQ} channel data sampled at 16,386 Hz. The fractal
dimension is evaluated with the VAR method and {\tt decimate}=64, which allows for the computation of $\fdim$ in real time. The top left panel shows one hour of data when the interferometer is in
its nominal observing mode ({\tt L1:DMT-ANALYSIS\_READY}) at the start of the glitchy period. Each point in the plot represents the value of the fractal dimension for a second-long data segment. The
fractal dimension is stationary with most values ranging between $\fdim=1$ (linear noise) and $\fdim=1.2$. The plot's shaded areas represent time segments where whistle glitches were identified and
flagged in the \ac{LVK} Data Quality Segment Database \cite{FISHER2021100677}. Three out of the four glitches flagged by the {\tt L1:DHC-WHISTLES} data quality flag coincide with anomalous values of
$\fdim>1.2$. The Q-transform \cite{Chatterji} plots (``Q-scans'') of these anomalous segments are shown in Fig.~\ref{QSCAN-L1_LSC-PRCL_OUT_DQ-var-dec_64-start_2020-02-04}. These Q-scans show that
all four time segments include whistle glitches. However, the glitch corresponding to the anomalous fractal dimension with value just above 1.2 (top-right panel) appears to be less loud than the
other three, possibly explaining why it is missed by the {\tt L1:DHC-WHISTLES} data quality flag. The corresponding Q-scans of the calibrated strain data used for astrophysical searches ({\tt
L1:DCS-CALIB\_STRAIN\_C01}) are shown in Fig.~\ref{QSCAN-DCS-CALIB_STRAIN_C01-var-dec_64-start_2020-02-04}. As expected, the fainter glitch is not visible in the Q-scan of the calibrated strain.

The top-right panel of Fig.~\ref{FD-L1_LSC-PRCL_OUT_DQ-var-dec_64-start_2020-02-04} shows the fractal dimension for one hour of later data which is characterized by a more severe, increasing rate of
whistle glitches. In the initial 45-minute stretch, the anomalous points of the fractal dimension show a clear correlation with the {\tt L1:DHC-WHISTLES} data quality flag segments. During this
time, the value of $\fdim$ becomes as high as $\sim 1.5$ denoting the appearance of louder glitches compared to the previous period. At about 15 minutes before the end of the segment, the glitch
rate and high \ac{SNR} render the interferometer data practically unusable for astrophysical searches. This is denoted with a new data quality flag, {\tt L1:DHC-SEVERE\_WHISTLES\_FEB4}. Eventually,
the excess noise leads to the interferometer dropping out of observing mode and losing lock. The bottom-left panel shows the fractal dimension for a one-hour of data during this time. The plot
starts with the detector initially not locked. Lock is regained at around $\sim 30$ minutes into the segment. The fractal dimension exhibits wild variations when the interferometer is unlocked with
$\fdim$ spanning the whole range of possible values before settling down again in the range $\sim 1$ - 1.2 when lock is regained. Note that the interferometer is still in non-observing mode during
this time while it is transitioning to its low-noise nominal state. Finally, the bottom-right panel shows $\fdim$ for a later one-hour of data when the instrument is again locked, in observing mode,
and free of whistle glitches. The fractal dimension is now stationary and well below $\fdim=1.2$. 

\begin{figure}[ht]
\begin{center}
\includegraphics[width=0.45\textwidth]{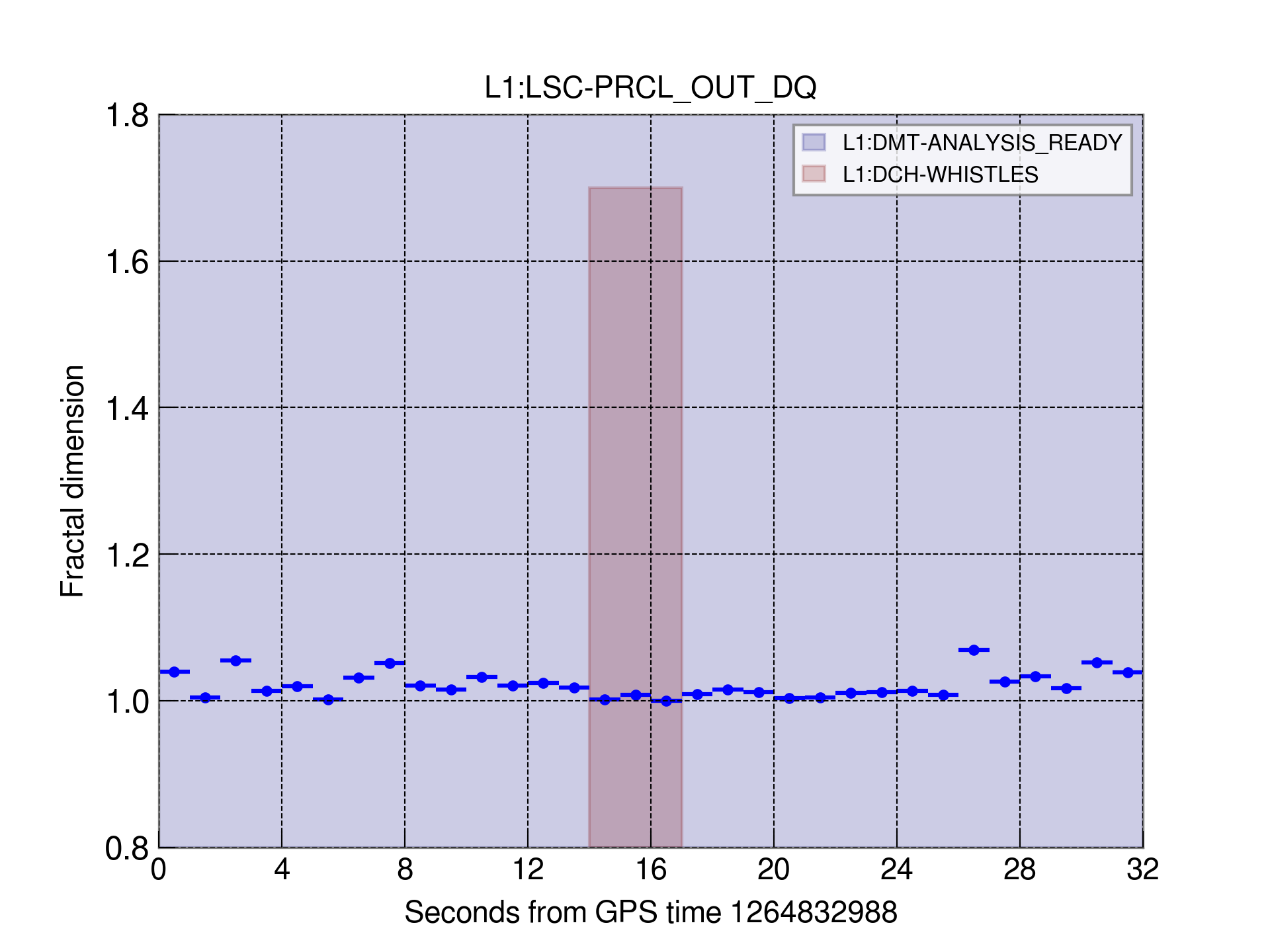}~
\raisebox{0.20\height}{\includegraphics[width=0.45\textwidth]{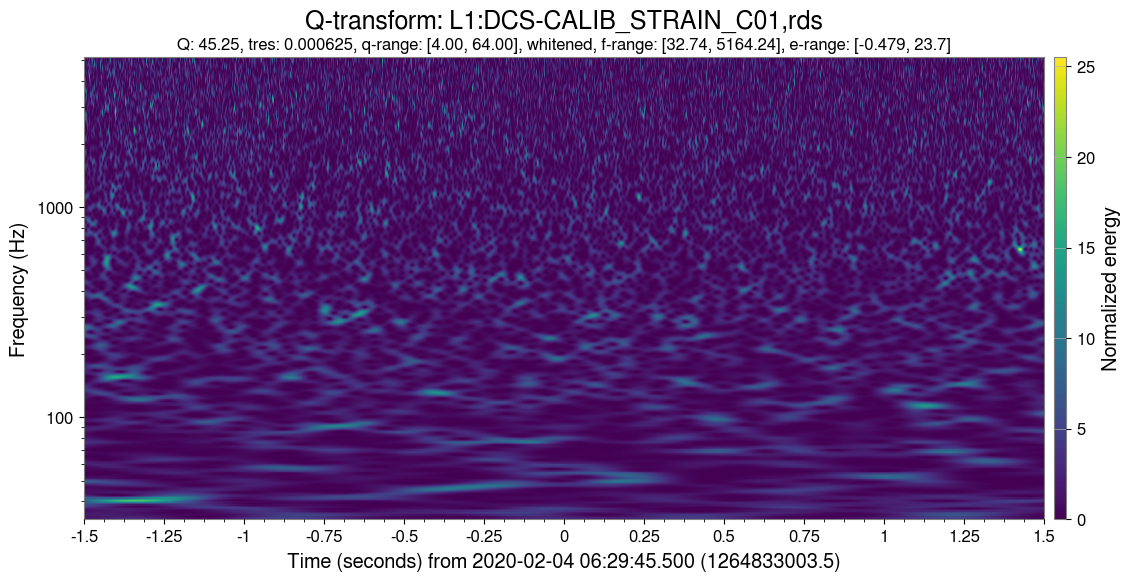}}\\
\includegraphics[width=0.45\textwidth]{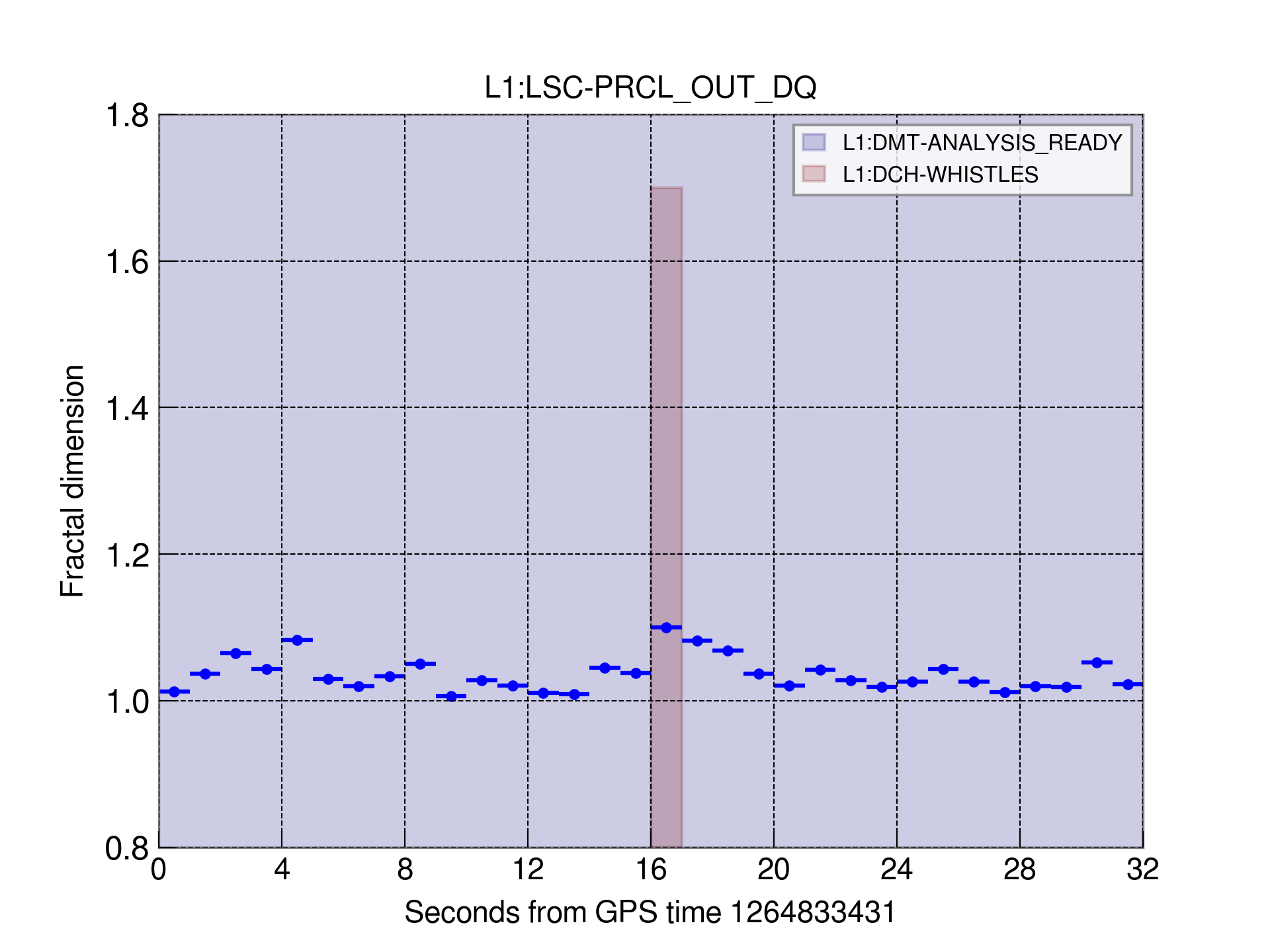}~
\raisebox{0.20\height}{\includegraphics[width=0.45\textwidth]{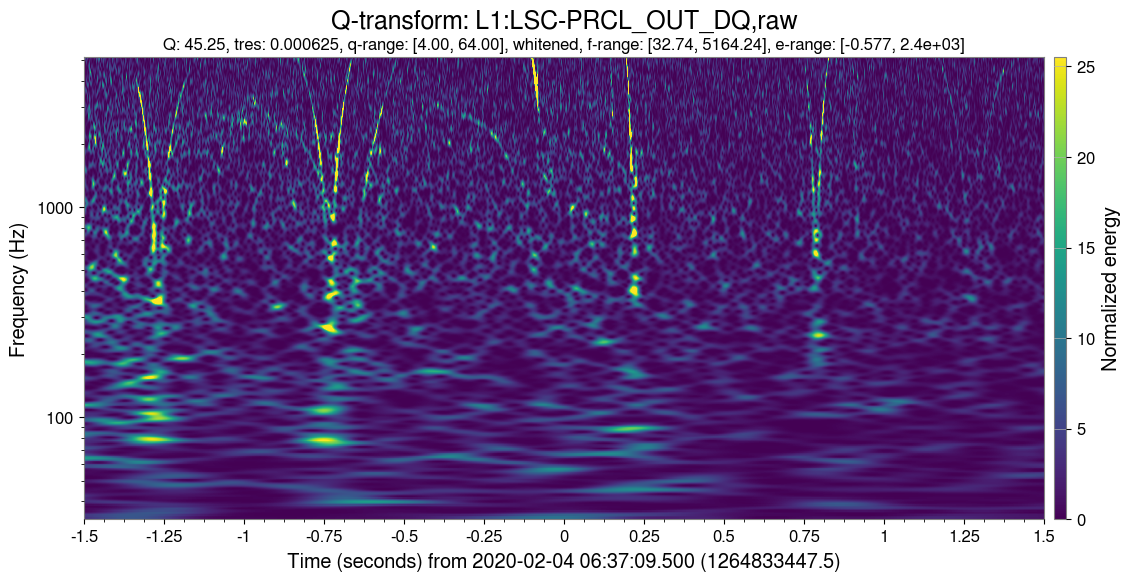}}
\caption{Variation of the fractal dimension and corresponding Q-scans around the two flagged segments in the top-left panel of Fig.\ \ref{FD-L1_LSC-PRCL_OUT_DQ-var-dec_64-start_2020-02-04} that do not appear to  be coincident with anomalous values of $\fdim$. The first period (GPS times 1264833002-1264833005) is quiet, thus the {\tt L1:DHC-WHISTLES} flag might have been incorrectly assigned. The second period (GPS time 1264833447) is characterized by whistles, although these are not as loud as the other flagged glitches in the one-hour period (see Fig.\ \ref{QSCAN-L1_LSC-PRCL_OUT_DQ-var-dec_64-start_2020-02-04}). In this case, the value of fractal dimension does not pass the $\fdim=1.2$ threshold. Additional, low-\ac{SNR} whistle glitches are present during these periods and can be identified by values of $\fdim$ approaching $1.1$.}
\label{FD-L1_LSC-PRCL_OUT_DQ-var-dec_64-start_2020-02-04-flagged}
\end{center}
\end{figure}

The one-hour period shown in the top-left panel also shows two additional segments flagged by {\tt L1:DHC-WHISTLES} where apparently the fractal dimension is not anomalous. It is worth digging a
little deeper into them. The left panels of Fig.\ \ref{FD-L1_LSC-PRCL_OUT_DQ-var-dec_64-start_2020-02-04-flagged} show the variation of the fractal dimension of {\tt L1:LSC-PRCL\_OUT\_DQ} for a
period of 32 seconds around these segments. Three second-long Q-scans centered on the flagged times are shown in the right panels. During the data quality segment starting at GPS time 1264833002,
the fractal dimension is around $\fdim=1$, a value typical of quiet times. In fact, the Q-scan of this segment does not show any visible excess noise and might have been wrongly flagged. (The
Q-scan of the calibrated strain channel at this time also does not show any visible glitch.) The data quality segment at GPS time 164833447 includes values of the fractal dimension as high as
$\fdim\sim 1.1$. The Q-scan to the right shows indeed that this segment contains whistle glitches, although not as loud as the other flagged whistle glitches in the one-hour period. It is
interesting to note that the fractal dimension around this time shows a ``wavy'' behavior with its value fluctuating between $\fdim\sim 1$ and $\fdim\sim 1.1$ in a quasi-periodic fashion. This
effect is due to the presence of whistle glitches across the entire period. For example, a Q-scan of the first few seconds of the 32-second interval in the bottom panel, where $\fdim$ shows a
``bump'', confirms that (low-\ac{SNR}) whistle glitches are also present at this time.

\begin{figure}[ht]
\begin{center}
\includegraphics[width=0.45\textwidth]{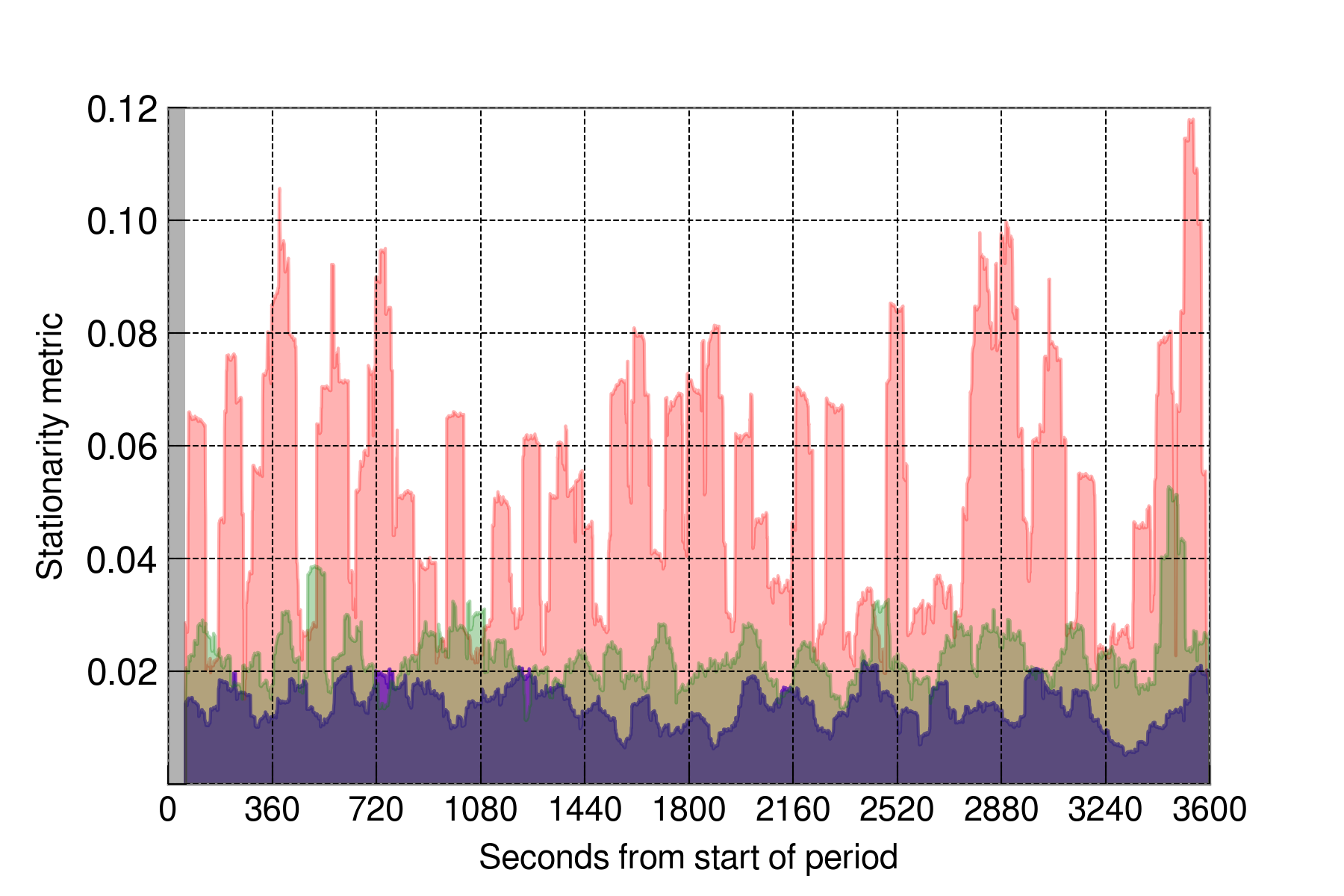}~
\includegraphics[width=0.45\textwidth]{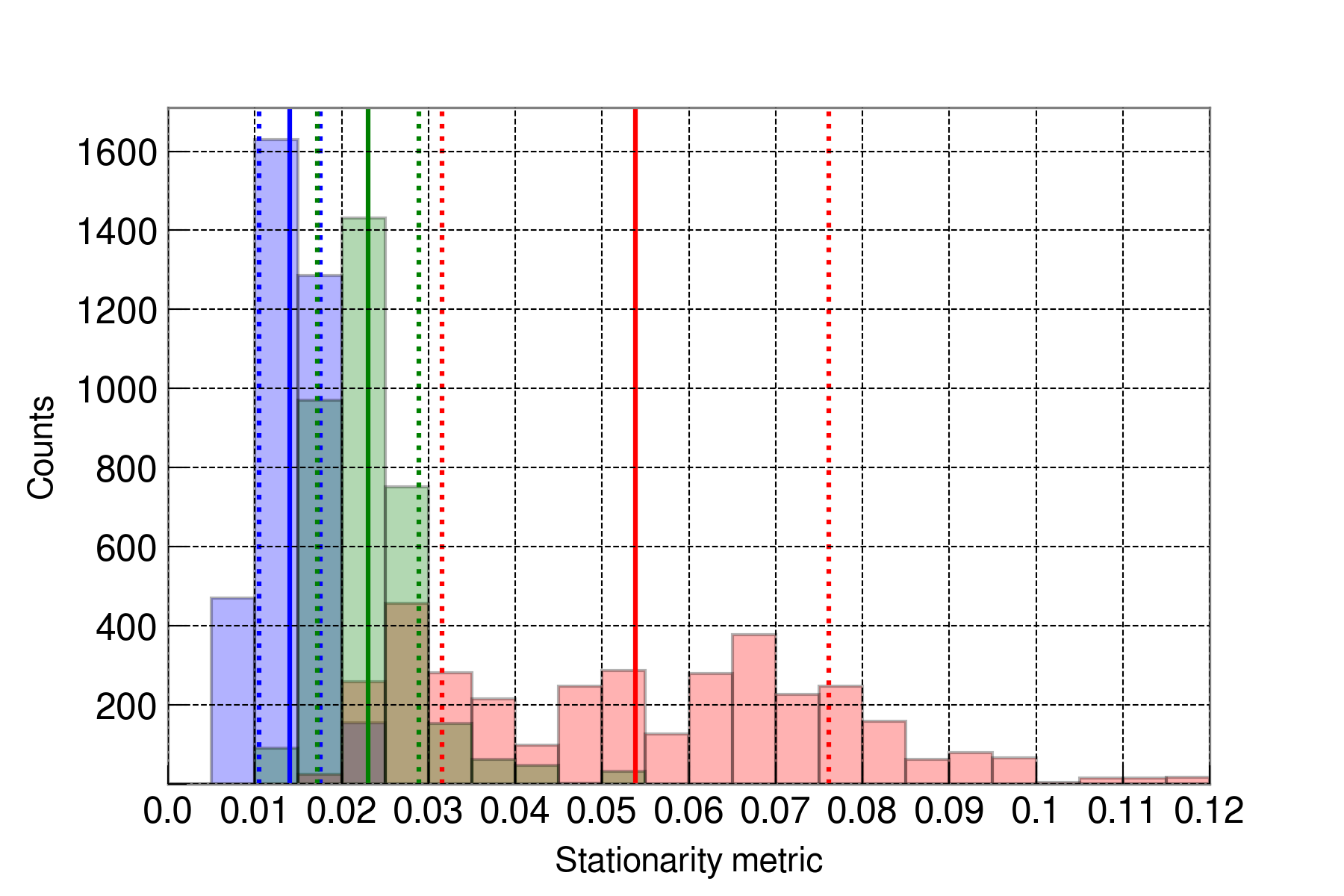}
\caption{Stationarity metric for the three observing mode data periods in Fig.\ \ref{FD-L1_LSC-PRCL_OUT_DQ-var-dec_64-start_2020-02-04} ({\tt L1:LSC-PRCL\_OUT\_DQ} channel). Left: Ratio of the rolling standard deviation of $\fdim$ to the rolling average of $\fdim$ computed on 60 second-long segments (darker-filled area: quiet time; medium dark-filled area: mild whistle glitchy time; darker-filled area: severe whistle glitchy time). The first 60 seconds of data (grey area) are used to generate the first value of the metric. Right: Histogram distribution of the stationarity metric for the three different time segments (blue: quiet time; green: mild glitchy time; red: severe glitchy time). Solid vertical lines indicate the mean values of the stationarity metric. Dotted vertical lines denote standard deviations from the means.}
\label{STAT-L1_LSC-PRCL_OUT_DQ-var-dec_64-start_2020-02-04}
\end{center}
\end{figure}

In addition to glitch identification, the fractal dimension can be used to characterize the (non-) stationarity of the interferometer in the presence of noise transients. It is straightforward to
note from Fig.~\ref{FD-L1_LSC-PRCL_OUT_DQ-var-dec_64-start_2020-02-04} that the variation of $\fdim$ is larger when whistle glitches occur than in their absence. This effect can be quantified by
computing the \emph{stationarity metric}, defined as ratio of the rolling standard deviation of $\fdim$ to the rolling average of $\fdim$. When glitches are present, the metric shows higher values
and vice versa. The metric can then be used to determine whether a given period is characterized by excess noise. The left panel of Fig.~\ref{STAT-L1_LSC-PRCL_OUT_DQ-var-dec_64-start_2020-02-04}
shows the stationarity metric for the three one-hour, observing periods of Fig.~\ref{FD-L1_LSC-PRCL_OUT_DQ-var-dec_64-start_2020-02-04} computed on a 60 second-long rolling window preceding each
value of $\fdim$. The quiet time (bottom-right panel of Fig.~\ref{FD-L1_LSC-PRCL_OUT_DQ-var-dec_64-start_2020-02-04} is characterized by a value of the stationarity metric consistently smaller than
0.02 (darker area). The moderate (Fig.~\ref{FD-L1_LSC-PRCL_OUT_DQ-var-dec_64-start_2020-02-04}, top-left) and the severe (Fig.~\ref{FD-L1_LSC-PRCL_OUT_DQ-var-dec_64-start_2020-02-04}, top-right
panel) periods exhibit higher values of the metric over their corresponding one-hour data stretches, with the severe period characterized by a value of the metric as high as $\sim 5$ times the value
for the moderate case. The right panel of Fig.~\ref{STAT-L1_LSC-PRCL_OUT_DQ-var-dec_64-start_2020-02-04} shows the histogram distributions of the stationarity metric for the three periods. A simple
threshold of $0.02$ captures all the 60-second segments with elevated excess noise.

To test the method on another class of noise we repeat the above analysis for the scattered light glitches. Excess noise transients due to unwanted scattered light in the interferometer optical
systems have been one of the major sources of noise in \ac{LIGO} data during \ac{O3} \cite{soni}. For that reason, a number of investigations to determine its causes and develop mitigation
strategies have been conducted over the past few years \cite{Soni_2021,Valdes:2017xce,Longo:2021avq,Bianchi:2021unp}. Scattered light noise manifests itself as transients at low and medium
frequencies in the interferometer sensitive band. It is commonly divided in two sub-classes: slow and fast scattering \cite{soni}.  Slow scattering typically occurs during periods of elevated
ground motion due to earthquakes and microseism noise in the sub-Hz frequency band. The time-frequency representation of slow scattered light glitches is that of stacked arches with duration of
few seconds. Fast scattering occurs at higher frequencies (a few Hz) due to anthropogenic noise.  Fast scattered light glitches have a smaller duration and \ac{SNR} than slow scattered light
glitches. Their time-frequency representation is that of arches with duration of $\sim 1$ second. (See Figs.\ 4 and 5 in Ref.\ \cite{Davis_2021} for Q-scans of scattering glitches.)

\begin{figure}[ht]
\begin{center}
\includegraphics[width=0.45\textwidth]{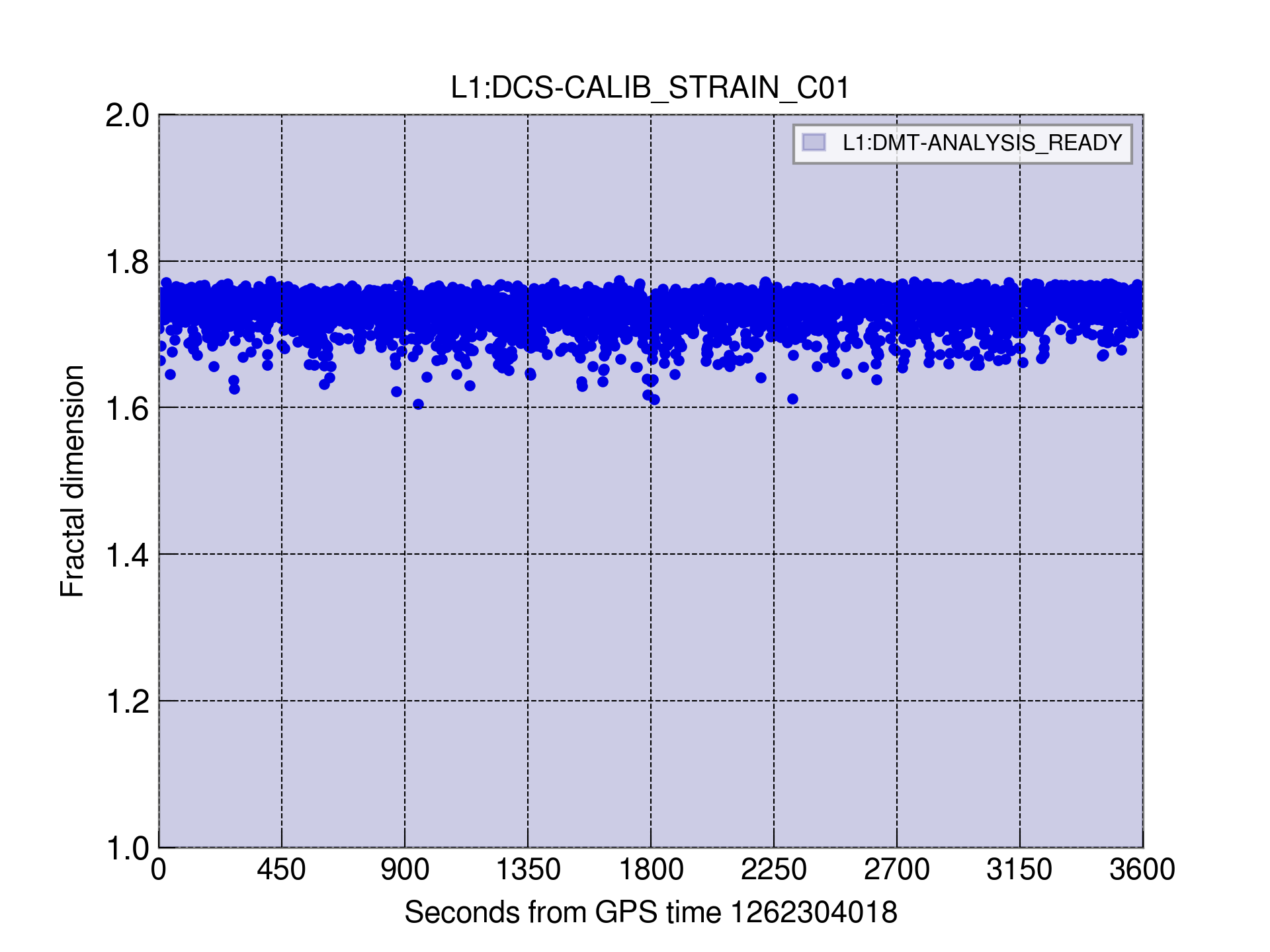}~
\includegraphics[width=0.45\textwidth]{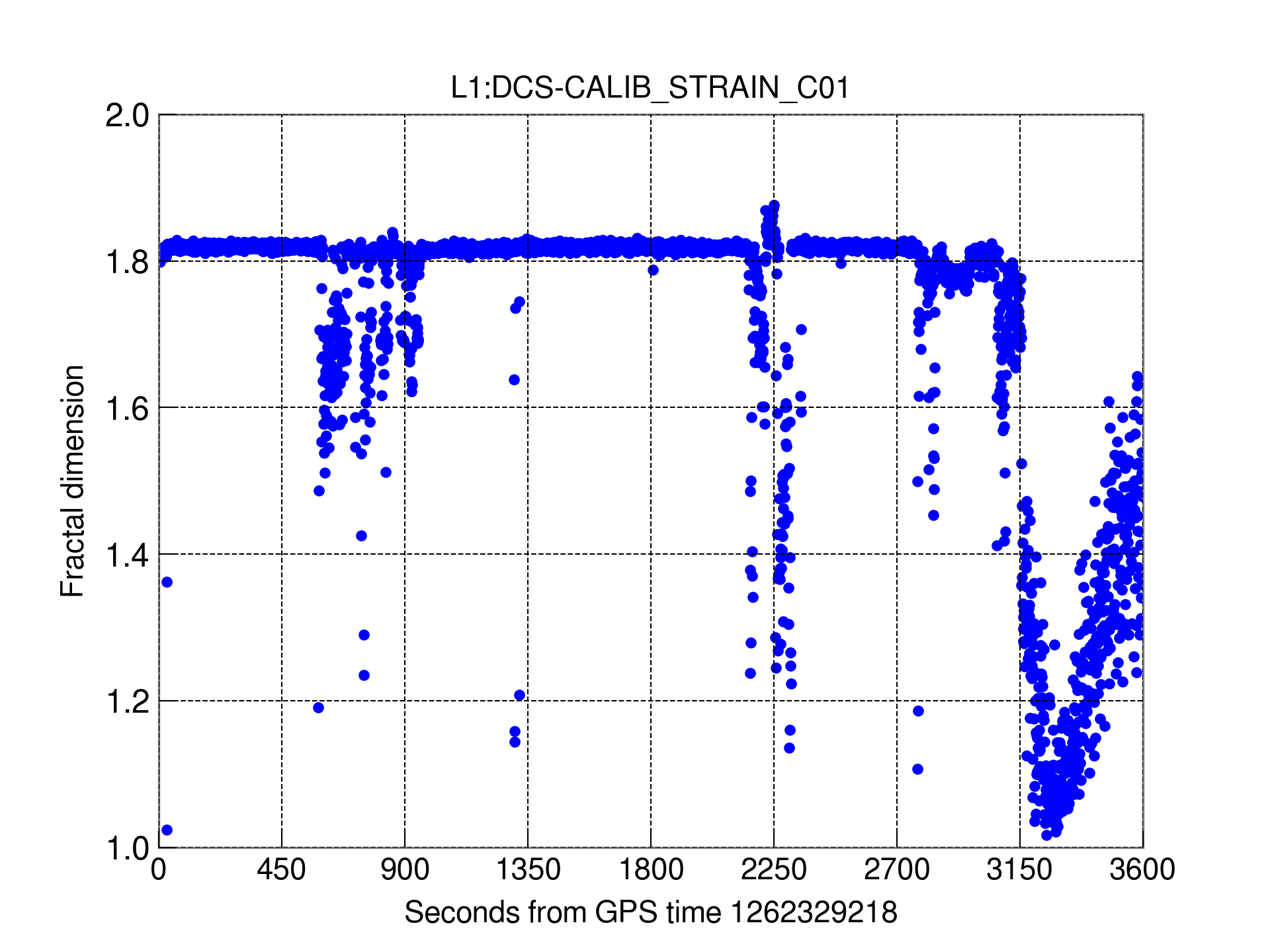}\\
\includegraphics[width=0.45\textwidth]{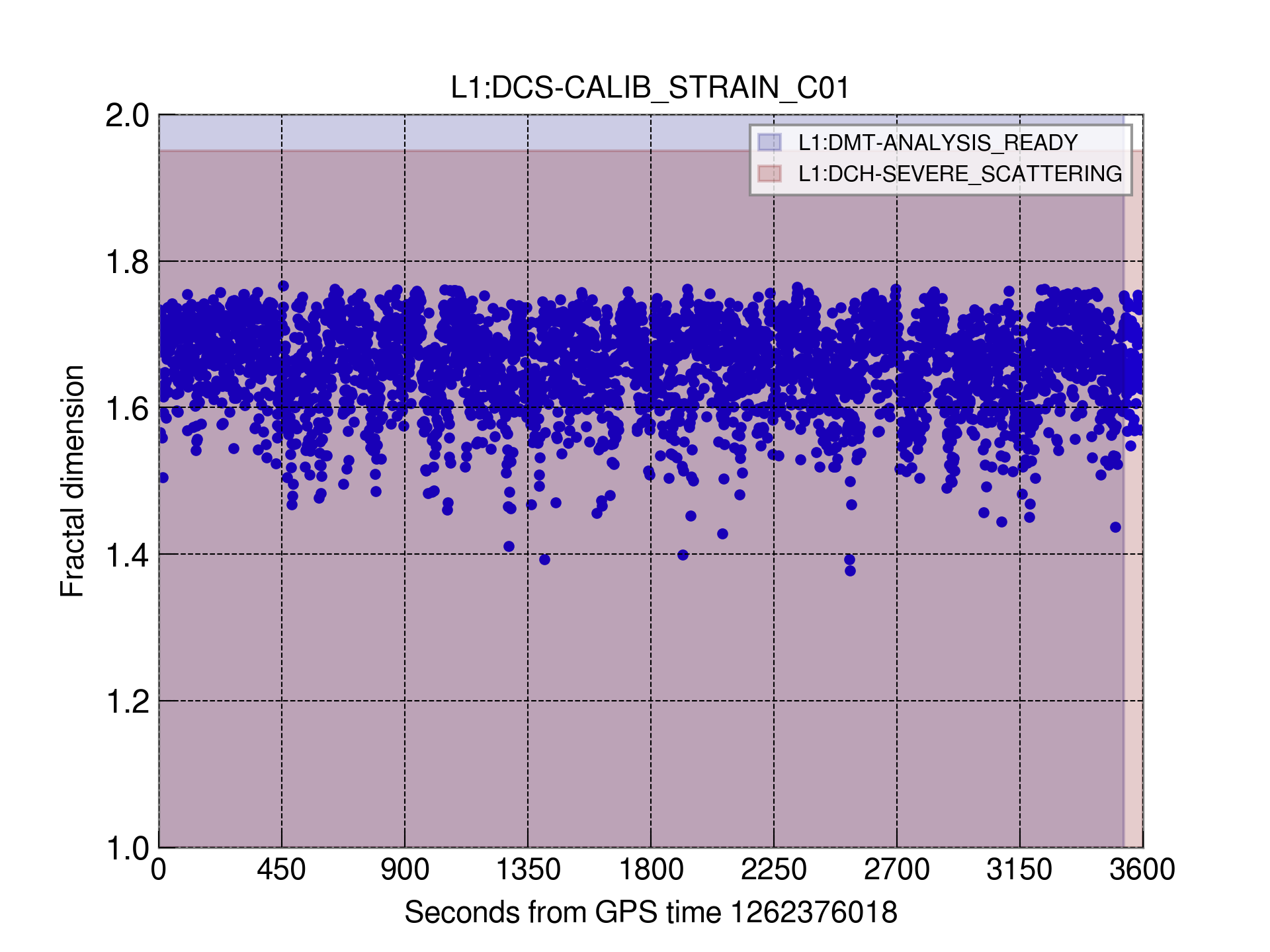}~
\includegraphics[width=0.45\textwidth]{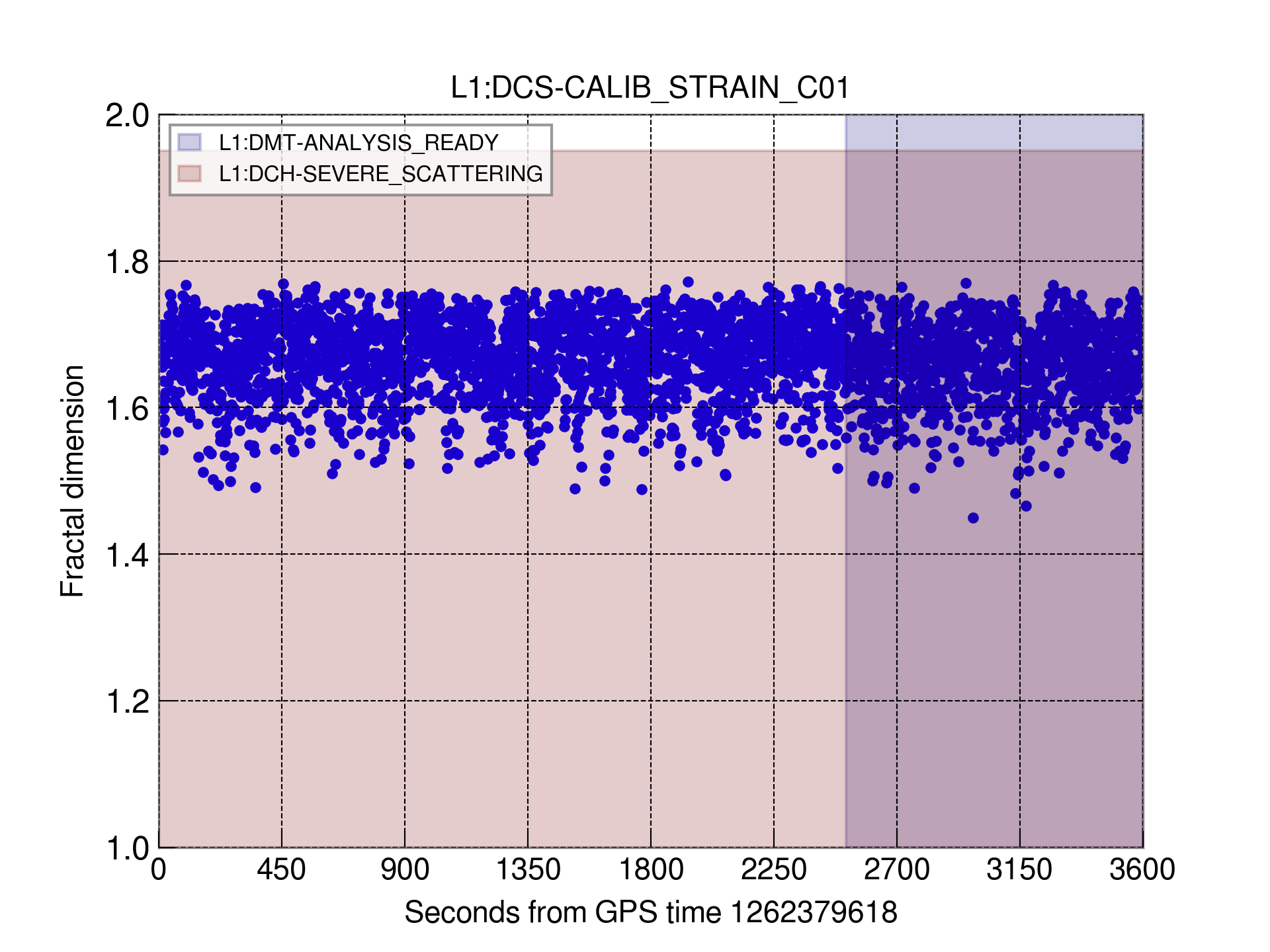}
\caption{Fractal dimension of the calibrated strain channel {\tt L1:DCS-CALIB\_STRAIN\_C01} for four, one-hour periods of \ac{L1} data with elevated scattered light glitch activity. The sampling rate of the channel is 16,384 Hz. The fractal dimension is computed with the VAR algorithm decimated at 64. Each point represents $\fdim$ for one second of data. The interferometer is in observing mode ({\tt L1:DMT-ANALYSIS\_READY}) during the entire period corresponding to the top-left panel, and out of observing mode for the entire period corresponding to the top-right panel and parts of the two bottom panel periods. The top-right panel corresponds to a period of high noise with several lock losses denoted by the sudden drops of $\fdim$. Different color shades denote periods of observing mode and active status of the {\tt L1:DHC-SEVERE\_SCATTERING} data quality flag.}
\label{FD-L1_DCS-CALIB_STRAIN_C01-var-dec_64-start_2020-01-06}
\end{center}
\end{figure}

\begin{figure}[ht]
\begin{center}
\includegraphics[width=0.45\textwidth]{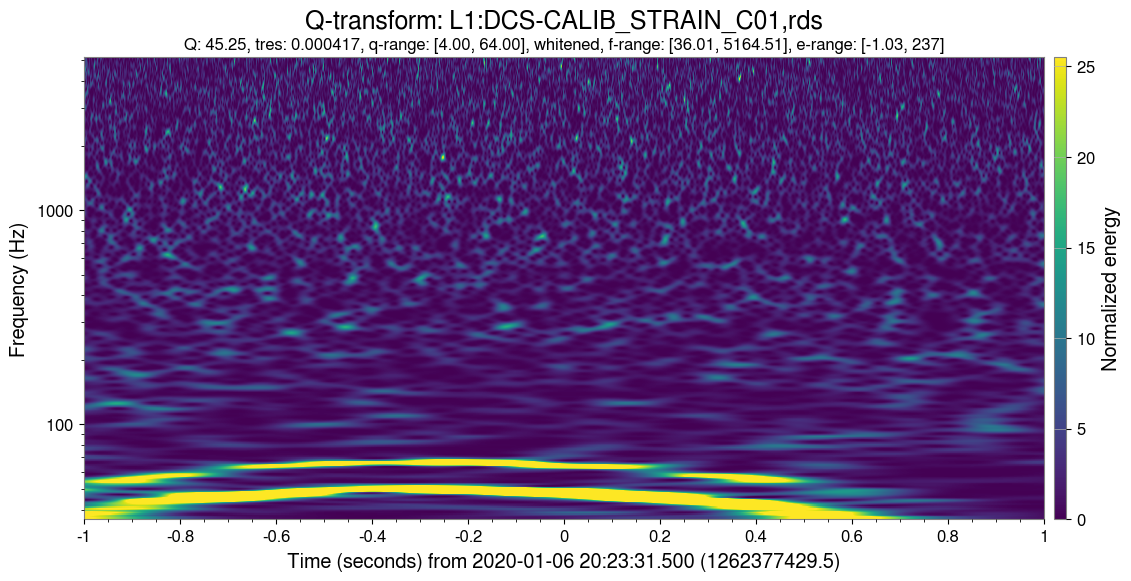}~
\includegraphics[width=0.45\textwidth]{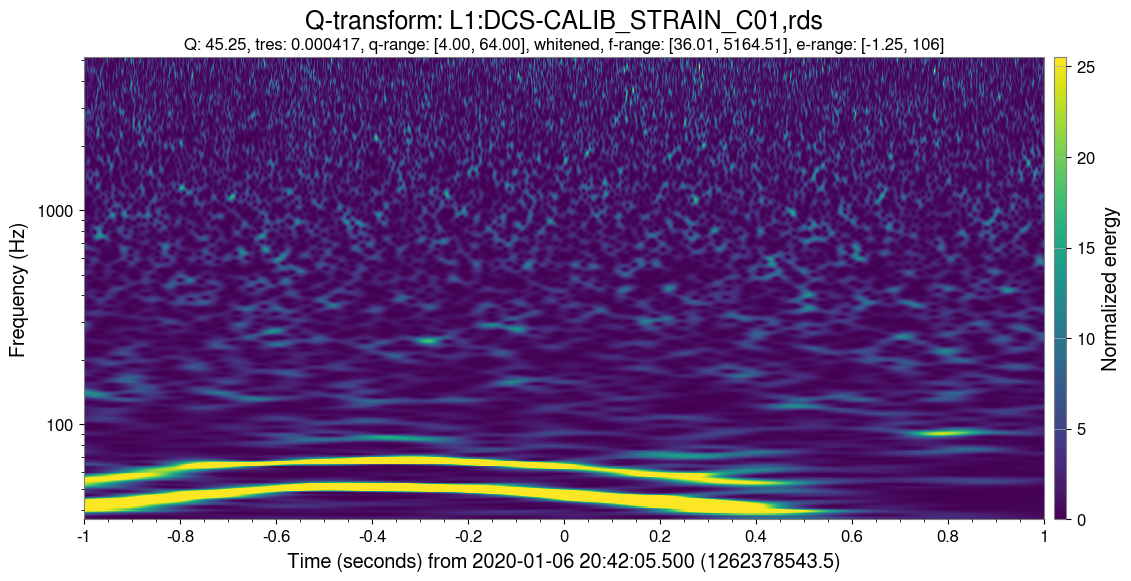}\\
\includegraphics[width=0.45\textwidth]{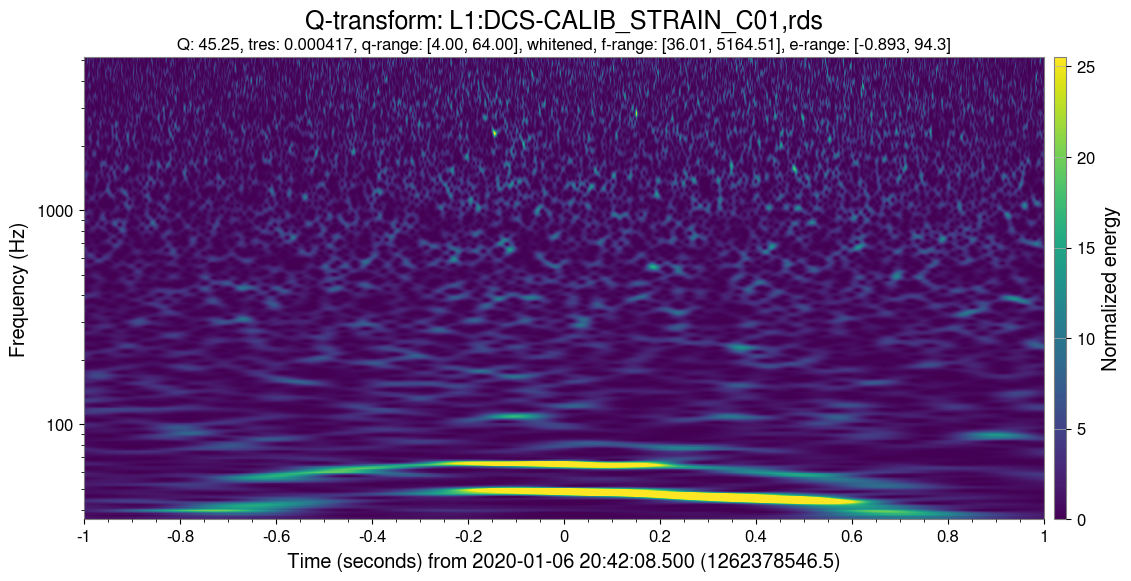}~
\includegraphics[width=0.45\textwidth]{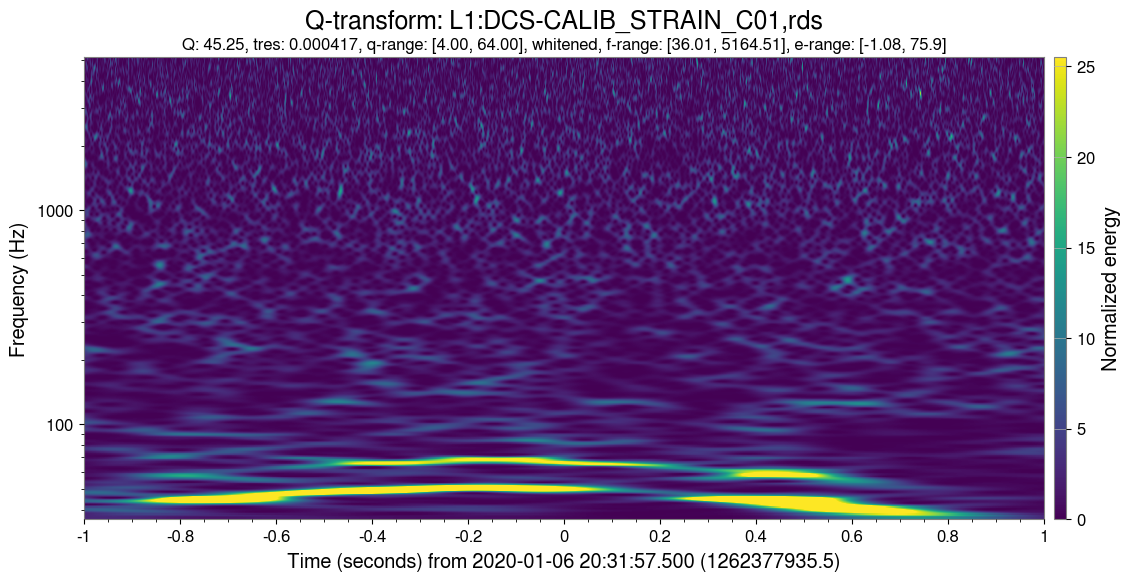}
\caption{Q-scans \cite{Chatterji} of the calibrated strain channel {\tt L1:DCS-CALIB\_STRAIN\_C01} for the four anomalous times with lowest value of $\fdim$ in the bottom-left of Fig.~\ref{FD-L1_DCS-CALIB_STRAIN_C01-var-dec_64-start_2020-01-06}. The times with anomalous fractal dimensions correspond to scattered light glitches \cite{Davis_2021}.}
\label{QSCAN-DCS-CALIB_STRAIN_C01-var-dec_64-start_2020-01-06}
\end{center}
\end{figure}

Figure \ref{FD-L1_DCS-CALIB_STRAIN_C01-var-dec_64-start_2020-01-06} shows the variation of the fractal dimension for four, one-hour periods of \ac{O3} calibrated strain \ac{L1} data on January 6, 2020, during a time of elevated slow scattered light glitch activity \cite{Soni_2021}. The top-left plot shows $\fdim$ before the onset of the glitchy period. The value of the fractal dimension is stationary over this period, taking values between $\fdim=1.6$ and 1.8 with occasional, low-\ac{SNR} isolated glitches identified by values of $\fdim\sim 1.6$. The two bottom plots show the value of the fractal dimension over two hours with elevated scattered light noise (denoted by an active {\tt L1:DHC-SEVERE\_SCATTERING} data quality flag) following a drop out of observing mode and a lock loss (top-right plot). The fractal dimension in the presence of scattered light glitches exhibits higher variability compared to the quiet time with values as low as $\fdim\sim 1.4$. Figure \ref{QSCAN-DCS-CALIB_STRAIN_C01-var-dec_64-start_2020-01-06} shows the Q-scans of the four anomalous times with lowest value of $\fdim$ in the bottom-left of Fig.~\ref{FD-L1_DCS-CALIB_STRAIN_C01-var-dec_64-start_2020-01-06}.

\begin{figure}[ht]
\begin{center}
\includegraphics[width=0.45\textwidth]{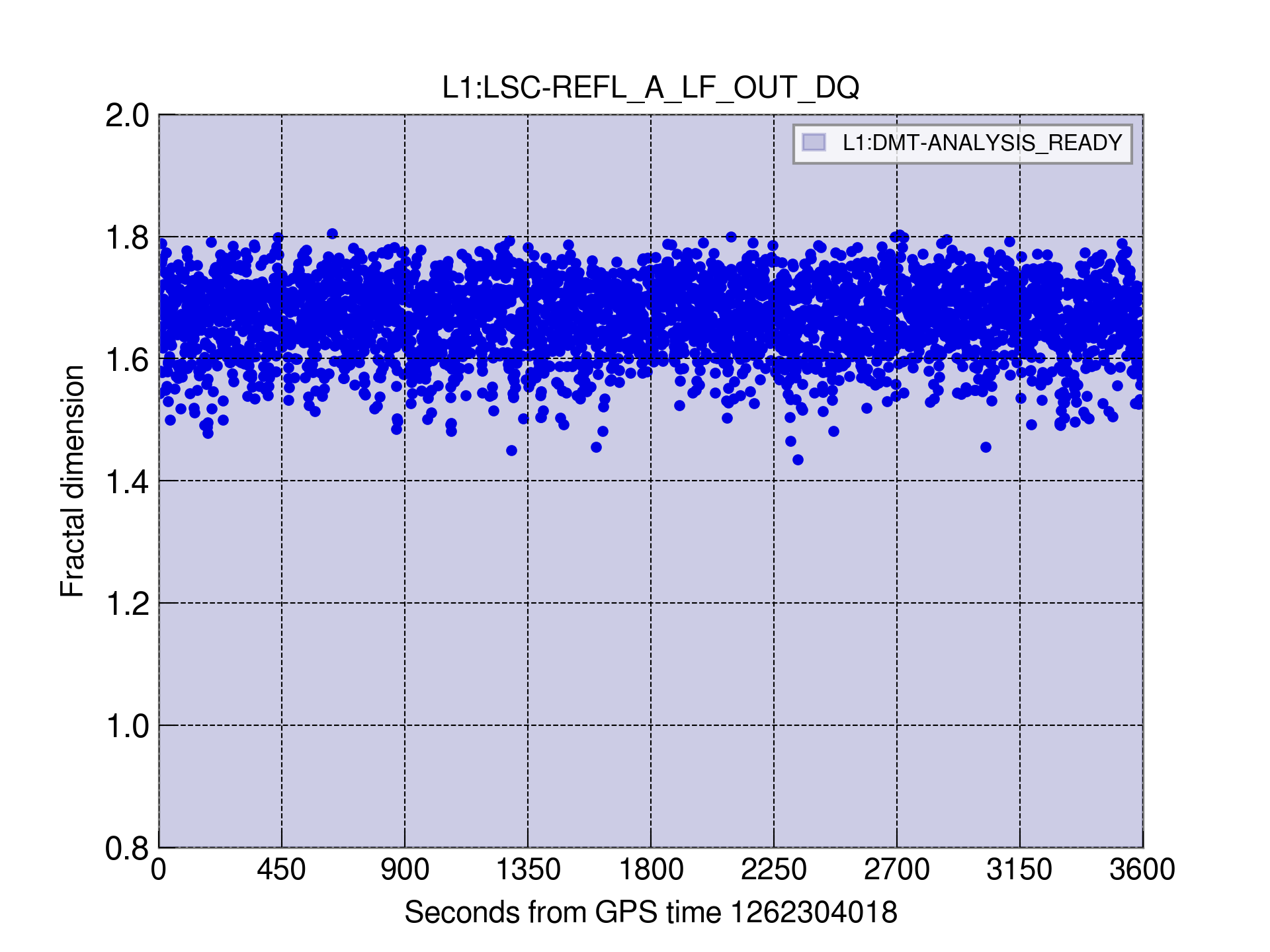}~
\includegraphics[width=0.45\textwidth]{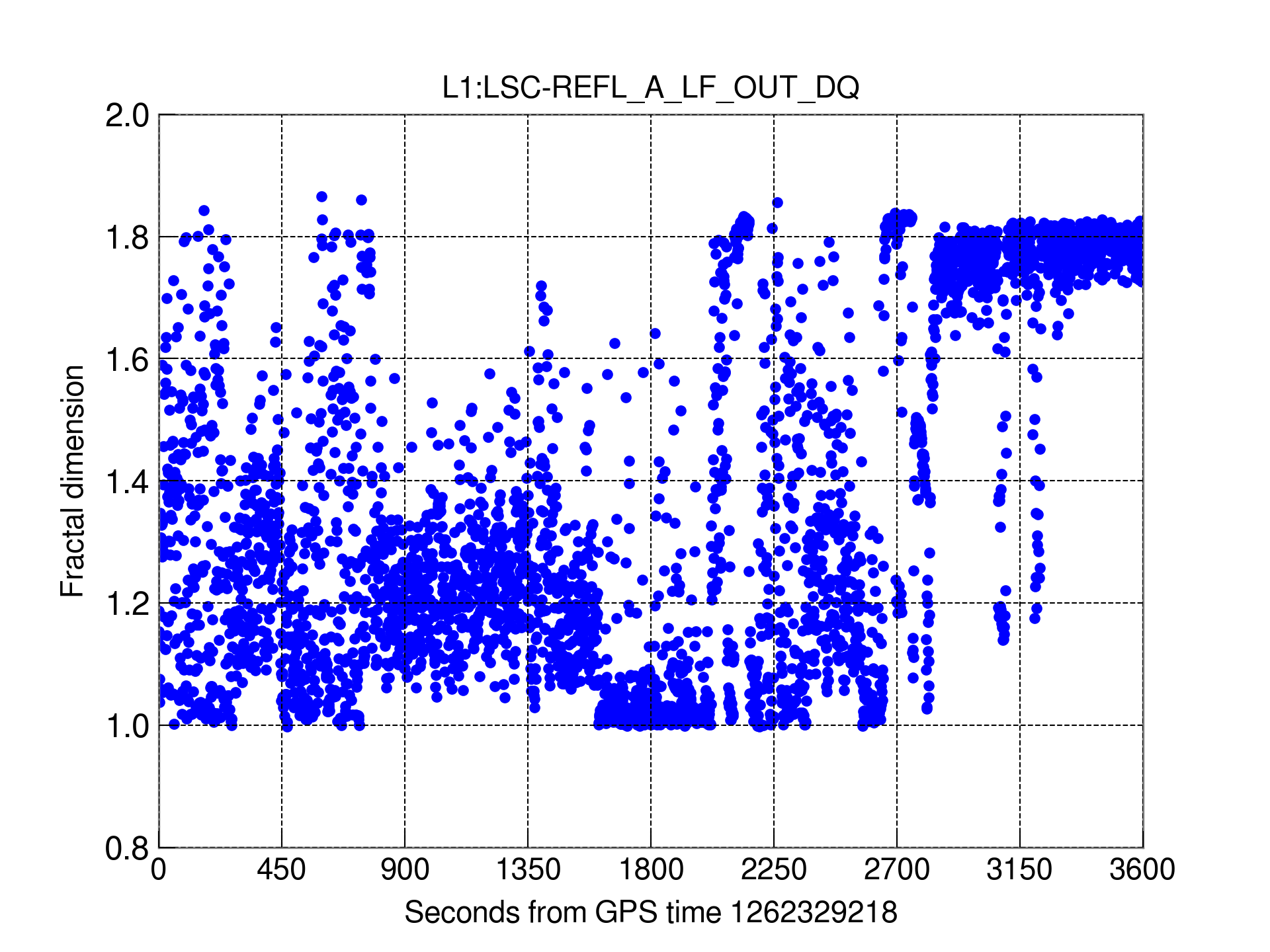}\\
\includegraphics[width=0.45\textwidth]{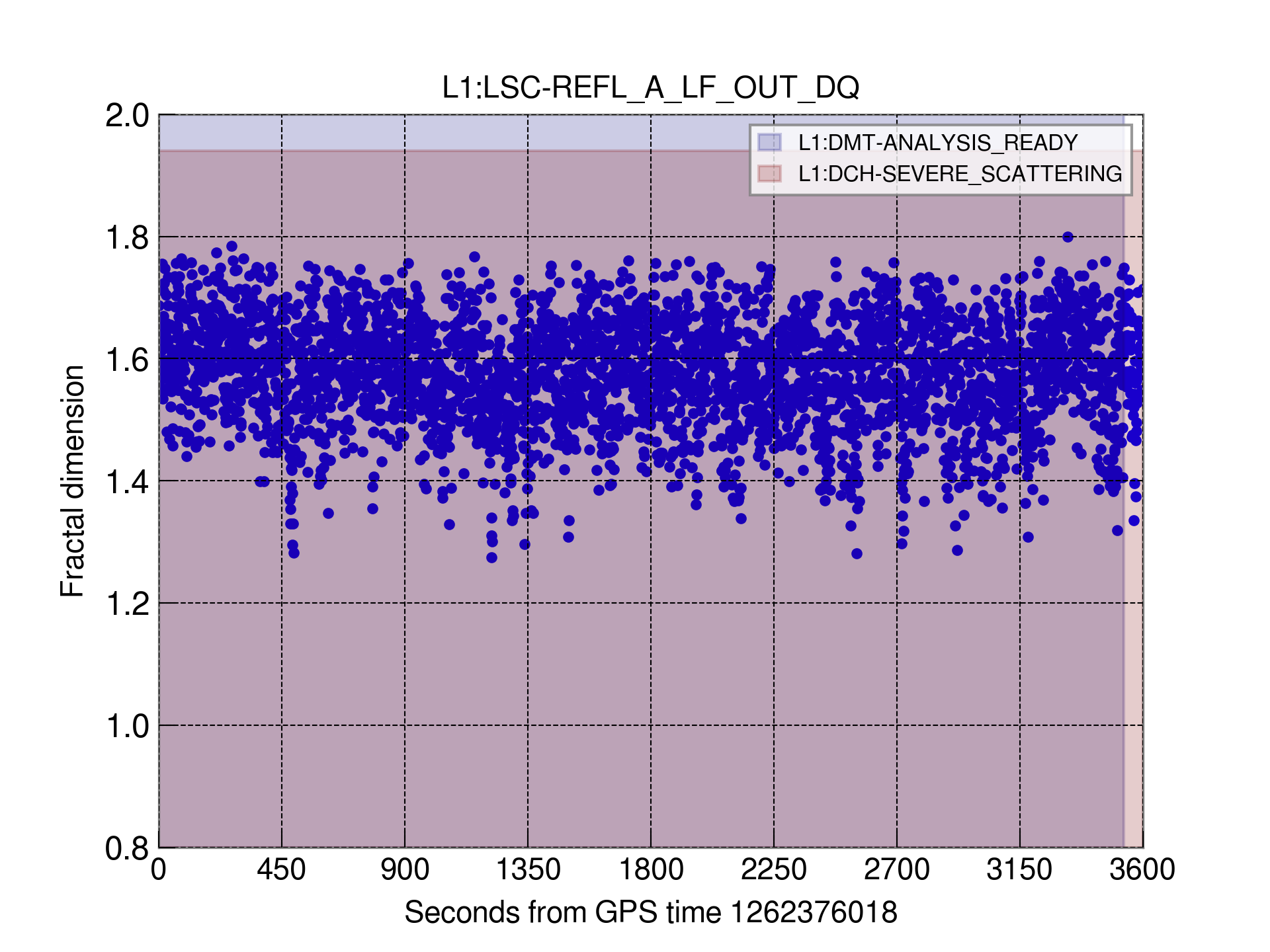}~
\includegraphics[width=0.45\textwidth]{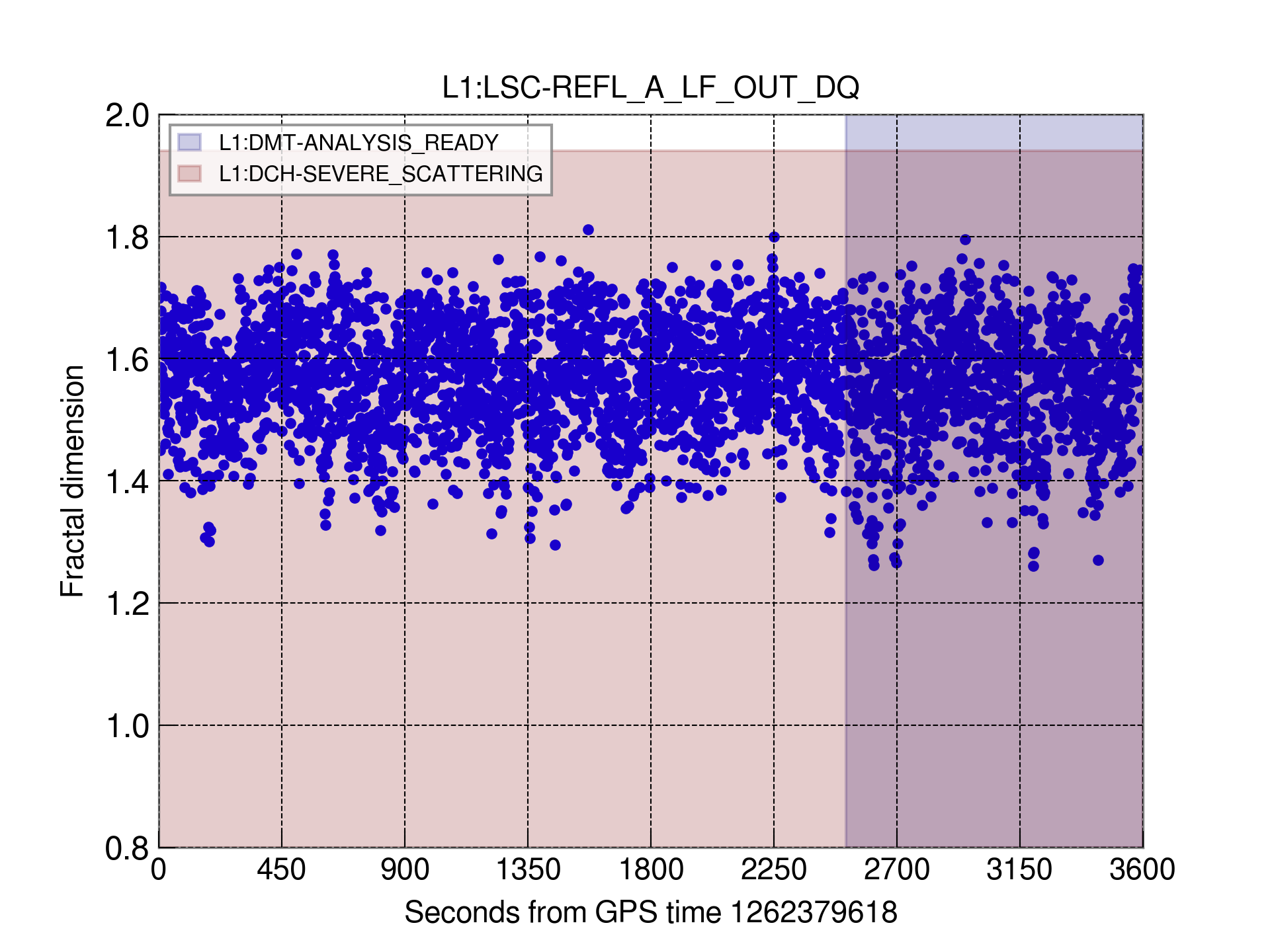}
\caption{Fractal dimension of the {\tt L1:LSC-REFL\_A\_LF\_OUT\_DQ} witness auxiliary channel for the four periods in Fig.~\ref{FD-L1_DCS-CALIB_STRAIN_C01-var-dec_64-start_2020-01-06}. The sampling rate of the channel is 16,384 Hz. The fractal dimension is computed with the VAR algorithm decimated at 64. The witness channel displays a higher variability than the calibrated strain channel, denoting a noiser {\tt L1:LSC-REFL\_A\_LF\_OUT\_DQ} data stream compared to {\tt L1:DCS-CALIB\_STRAIN\_C01}.}
\label{FD-L1_LSC-REFL_A_LF_OUT_DQ-var-dec_64-start_2020-01-06}
\end{center}
\end{figure}

\begin{figure}[ht]
\begin{center}
\includegraphics[width=0.45\textwidth]{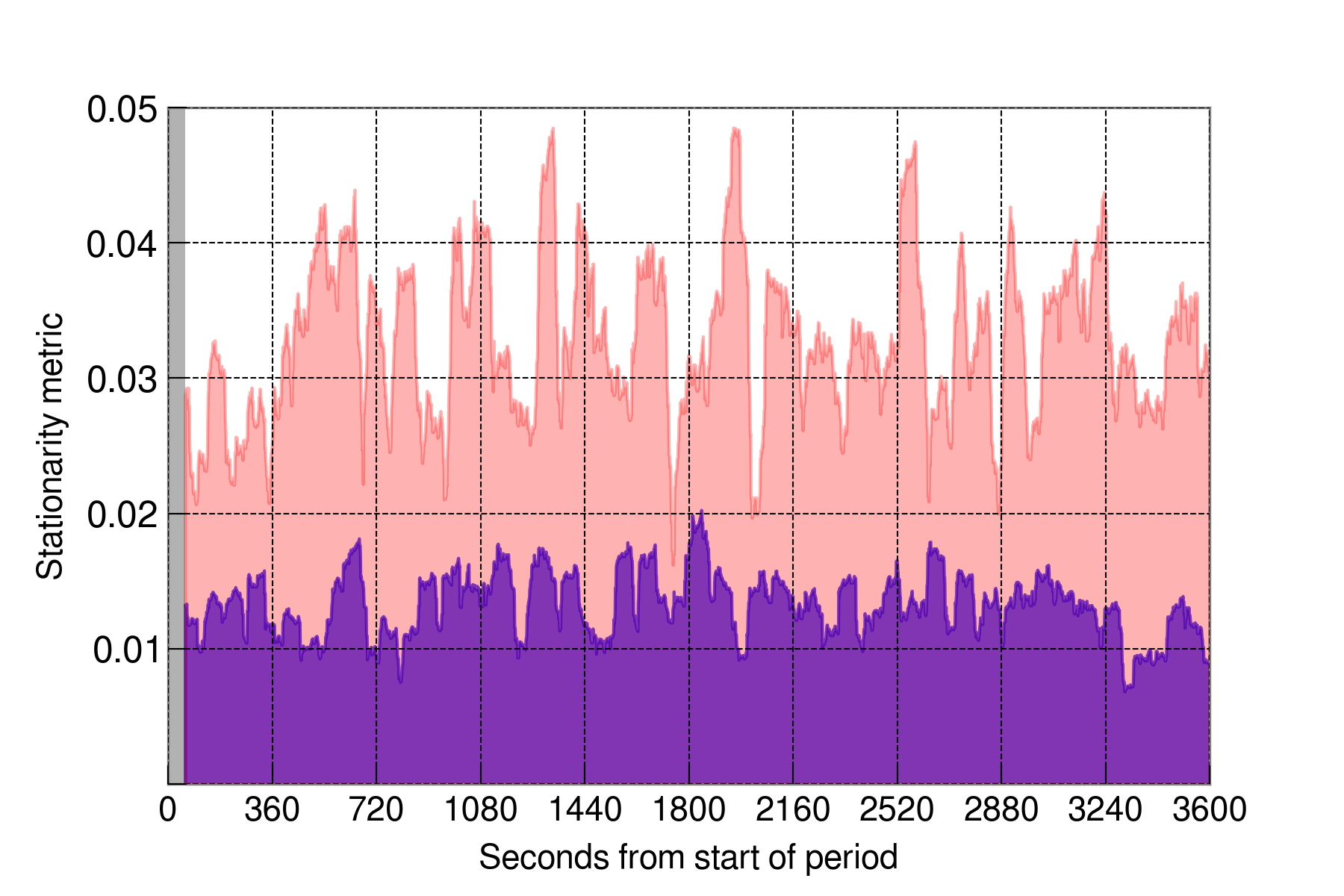}~
\includegraphics[width=0.45\textwidth]{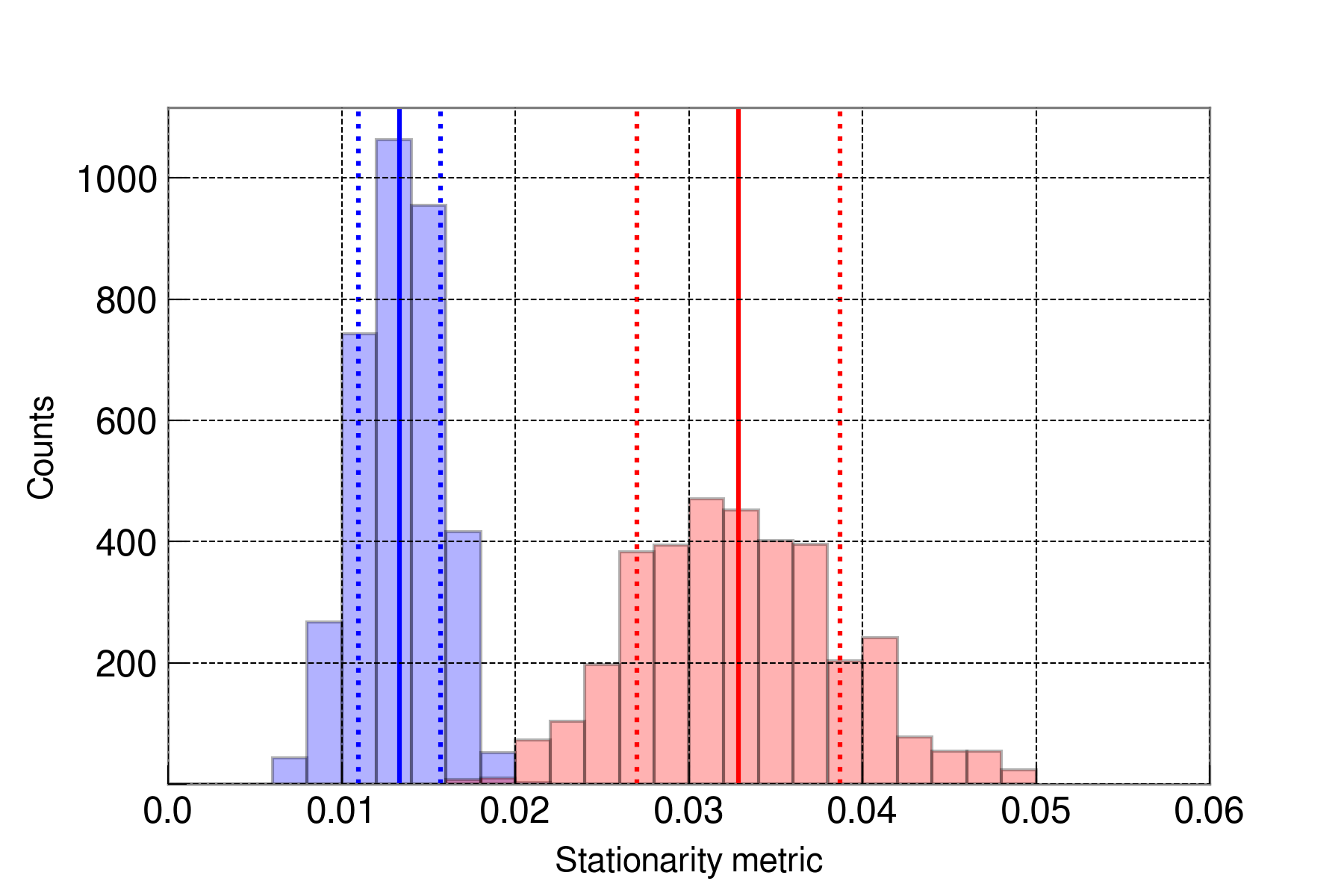}
\caption{Stationarity metric for the fractal dimension corresponding to the left panels of Fig.~\ref{FD-L1_DCS-CALIB_STRAIN_C01-var-dec_64-start_2020-01-06} ({\tt L1:DCS-CALIB\_STRAIN\_C01}
channel). Left: Ratio of the rolling standard deviation to the rolling average of $\fdim$ computed on 60 second-long segments (darker-filled area: quiet time, light-filled area: scattered light
glitchy time). The first 60 seconds of data (grey area) are used to generate the first value of the metric. Right: Histogram distributions of the stationarity metric for the two time periods (blue:
quiet time; red: glitchy time). Solid vertical lines indicate the mean values of the stationarity metric. Dotted vertical lines denote standard deviations from the means.}
\label{STAT-L1_DCS-CALIB_STRAIN_C01-var-dec_64-start_2020-01-06}
\end{center}
\end{figure}

\begin{figure}[ht]
\begin{center}
\includegraphics[width=0.45\textwidth]{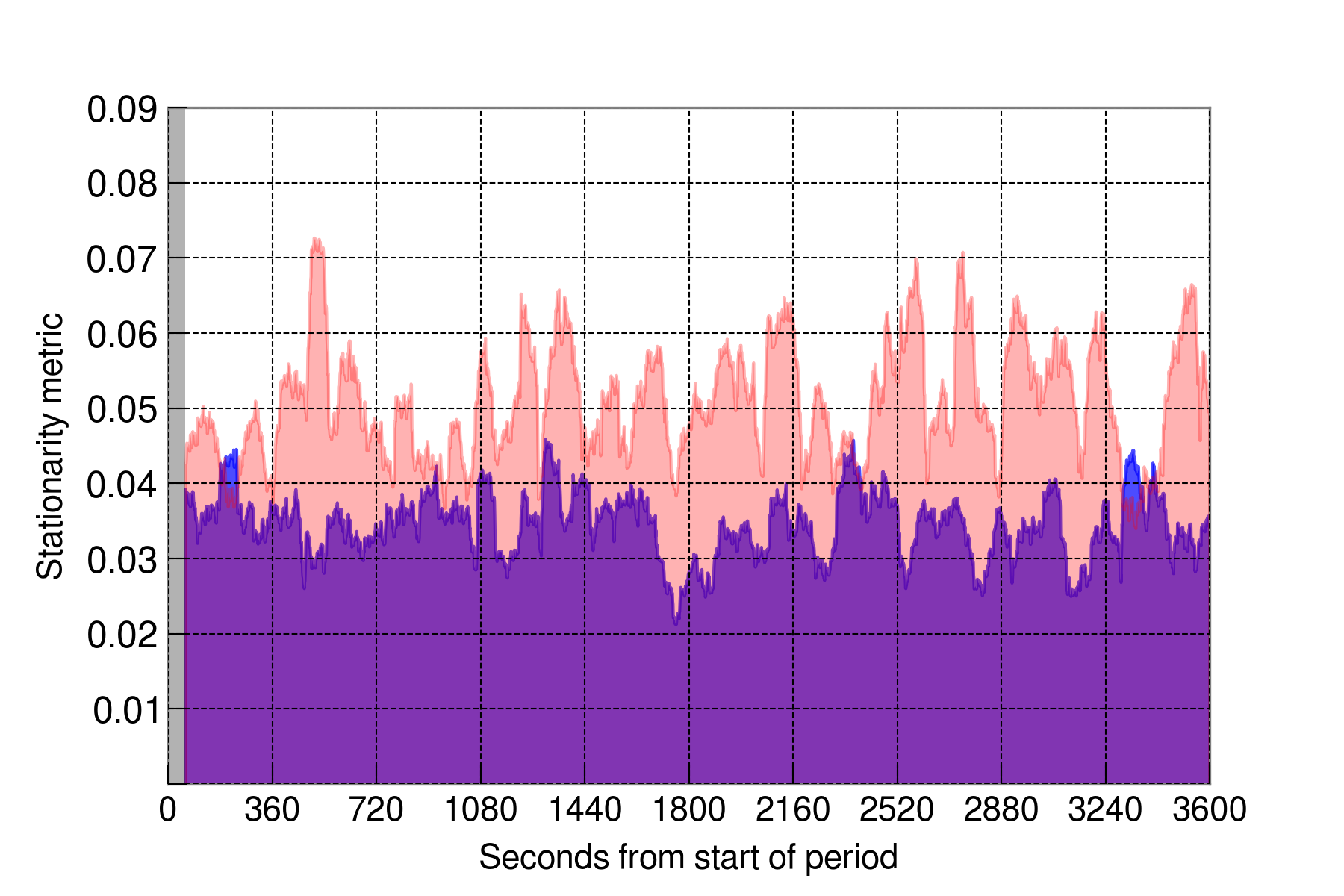}~
\includegraphics[width=0.45\textwidth]{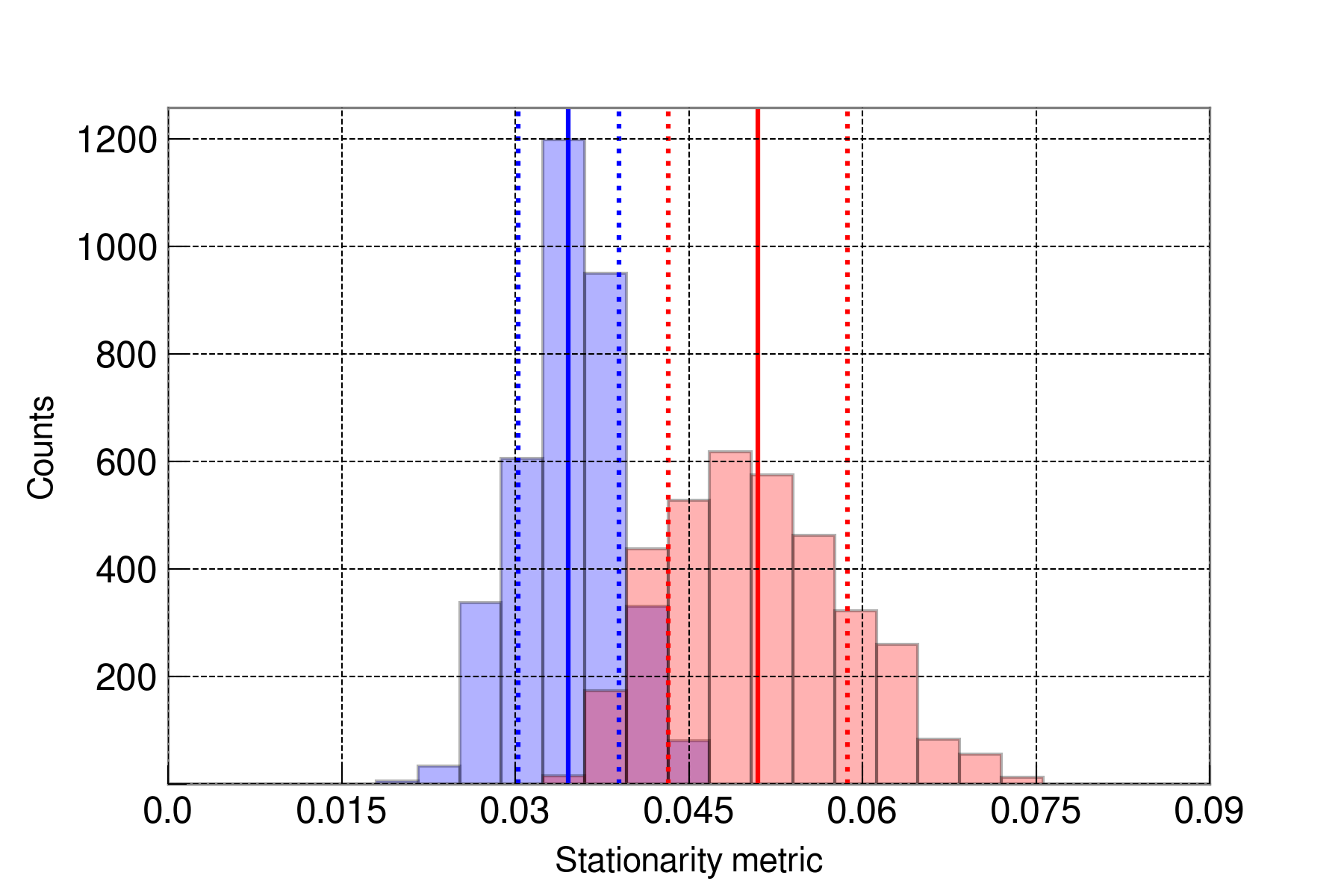}
\caption{Stationarity metric for the fractal dimension corresponding to the left panels of Fig.~\ref{FD-L1_LSC-REFL_A_LF_OUT_DQ-var-dec_64-start_2020-01-06} ({\tt L1:LSC-REFL\_A\_LF\_OUT\_DQ}
channel). Left: Ratio of the rolling standard deviation to the rolling average of $\fdim$ computed on 60 second-long segments (darker-filled area: quiet time, light-filled area: glitchy time). The
first 60 seconds of data (grey area) are used to generate the first value of the metric. Right: Histogram distributions of the stationarity metric for the two time periods (blue: quiet time; red:
scattered light glitchy time). Solid vertical lines indicate the mean values of the stationarity metric. Dotted vertical lines denote standard deviations from the means.}
\label{STAT-L1_LSC-REFL_A_LF_OUT_DQ-var-dec_64-start_2020-01-06}
\end{center}
\end{figure}

Figure \ref{FD-L1_LSC-REFL_A_LF_OUT_DQ-var-dec_64-start_2020-01-06} shows the variation of the fractal dimension of the length sensing and control auxiliary channel recording the output of the
photodiode which observes the reflected light from the \ac{PRC}, {\tt L1:LSC-REFL\_A\_LF\_OUT\_DQ}, for the same periods. The presence of scattered light excess noise manifests itself as an
increased variability of $\fdim$ in the witness channel. Similarly to the whistle glitch analysis, the excess noise can be quantified by comparing the stationarity metric in the different time
periods. Figures \ref{STAT-L1_DCS-CALIB_STRAIN_C01-var-dec_64-start_2020-01-06} and \ref{STAT-L1_LSC-REFL_A_LF_OUT_DQ-var-dec_64-start_2020-01-06} show the variation of the metric and its histogram
distribution for the two left panels of the calibrated strain and the {\tt L1:LSC-REFL\_A\_LF\_OUT\_DQ} auxiliary channel, respectively. The stationarity metric during glitchy times (lighter red
shaded area) is on average higher by a factor $\times 2$ compared to the metric during the quiet period (darker blue area) with the witness channel displaying an overall higher variability than the
strain channel. A simple threshold of 0.02 (0.04) on the stationarity metric for the strain (auxiliary) channel can be used to distinguish low- and high-noise periods.

\section{Conclusions} \label{conclusions}

Fractal analysis provides an effective method for characterizing the output of a physical measuring device. In this paper, we have applied this concept to \ac{GW} detector data. The fractal
dimension of the interferometer's strain and auxiliary channels can be used to identify noise artifacts in the data, monitor the stationarity of the detector and its status, as well as provide a
measure of data quality. To illustrate the method we have considered two known examples of noise transients that occurred during the latest \ac{LVK} observing run in one of the \ac{LIGO}
detectors (\ac{L1}). These examples show that the fractal dimension $\fdim$ is stationary and Gaussian distributed when the interferometer is operating in its nominal, low-noise mode.
Non-astrophyscal noise transients correlate with anomalous values of $\fdim$. Periods of elevated glitchiness are characterized by an increased variability of the fractal dimension and/or
non-stationarity. They can be identified by comparing the degree of dispersion of $\fdim$ with respect to quiet times. Periods of lock loss are denoted by extreme variations of the fractal
dimension over the whole range of possible values. 

Algorithms for the calculation of the fractal dimension are computationally cheap. One second of \ac{LIGO} data channel at 16,384 Hz can be processed with a numba \cite{numba} decorator on a single
GPU in $\sim 0.6$ seconds. Thus the method can be applied to \ac{GW} data analysis in real time. Identification of glitch times through identification of anomalous fractal dimension outliers is a
safe procedure against vetoing astrophysical \ac{CBC} signals. \ac{CBC} signals are characterized by well-defined time-frequency relations, i.e., in a sense they are ``smooth'' functions. The value
of the fractal dimension depends on the whole frequency content of the data. Therefore, adding a smooth function to the background noise is not expected to change the fractal dimension. This
conclusion may be different in the case of broadband, stochastic signals such as \ac{GW} from core-collapse supernovae. Whether the fractal dimension could be used to detect these signals is an
intriguing possibility which remains to be seen. 

Other avenues for future investigations on the applicability of the method to real data include development of anomaly detection algorithms in real-time (see, e.g, the interesting proposal in Ref.\
\cite{Ding:2021ekq}), a full study of the relation between the fractal dimension and the characteristics of the instrument noise (non-linearity effects, multi-band analysis, correlations between the
fractal dimension of the strain and auxiliary channels, relation to glitch \ac{SNR}, rate, frequency breadth or other features,\dots), and a more extensive characterization of real data (follow-up
of anomalous fractal dimension times, fractal dimension-based differentiation of glitch classes, construction of suitable data quality metrics, \dots). Theoretical developments include exploring
different analytical models for the definition of the fractal dimension, such as entropic models \cite{10.1016/j.camwa.2013.01.017,e21080720}, and other choices of numerical approximants for even
faster and more accurate estimations of $\fdim$.

\section*{Acknowledgements}

This work was partially supported by the U.S.\ National Science Foundation award PHY-2011334 and is based upon work supported by NSF's LIGO Laboratory which is a major facility fully funded by the
National Science Foundation. The author is grateful for computational resources provided by the LIGO Laboratory and supported by the U.S.\ National Science Foundation Grants PHY-0757058 and
PHY-0823459, as well as resources from the Gravitational Wave Open Science Center, a service of the LIGO Laboratory, the LIGO Scientific Collaboration and the Virgo Collaboration. Part of this
research has made use of data, software and web tools obtained from the Gravitational Wave Open Science Center \cite{gwosc} and publicly available at \url{https://www.gw-openscience.org}.

LIGO was constructed and is operated by the California Institute of Technology and Massachusetts Institute of Technology with funding from the U.S.\ National Science Foundation under grant PHY-0757058. Virgo is funded by the French Centre National de la Recherche Scientifique (CNRS), the Italian Istituto Nazionale di Fisica Nucleare (INFN) and the Dutch Nikhef, with contributions by Polish and Hungarian institutes. The author would like to thank the many colleagues of the LIGO Scientific Collaboration and the Virgo Collaboration who have provided invaluable help over the years and useful comments about this work, in particular Beverly Berger and Julian Ding.

Part of this work was done over the fall 2021 at the Institute of Pure and Applied Mathematics (IPAM), University of California Los-Angeles (UCLA). The author would like to thank IPAM, UCLA and the National Science Foundation through grant DMS-1925919 for hosting him. The author would also like to thank, in particular, Dima Shlyakhtenko for useful suggestions about the theoretical aspects of this work, and Christian Ratsch, Selenne Ba\~nuelos and all IPAM staff for their warm hospitality.

Software for this analysis is written in Python 3.x \cite{10.5555/1593511} (\url{https://www.python.org/}) and uses standard Open Source libraries and community-contributed modules from the Python
Package Index (PyPI) repository \cite{pypi} including numpy, scipy, gwpy, h5py, pandas, matplotlib, nds2utils, numba \cite{numba}, and sklearn \cite{scikit-learn}. Q-transform \cite{Chatterji} plots
have been generated with Ligo DV-Web \cite{Areeda:2016mee}. 

This manuscript has been assigned LIGO Document Control Center number LIGO-P2200005.

\section*{References}
\bibliographystyle{apsrev}
\bibliography{bibliography}

\end{document}